\documentclass[smallextended]{svjour3}                     
\smartqed  
\usepackage{amsmath}
\usepackage{bm}
\usepackage{graphicx}
\usepackage[authoryear]{natbib}
\usepackage{algorithm}
\usepackage{algpseudocode}
\usepackage{subfigure}
\usepackage{lineno,setspace}
\usepackage{setspace,xspace}
\usepackage{url}

\usepackage{ifpdf}
\usepackage{graphicx,natbib,amssymb,lineno}
\ifpdf
\usepackage[%
  pdftitle={Instructions for use of the document class
    elsart},%
  pdfauthor={Simon Pepping},%
  pdfsubject={The preprint document class elsart},%
  pdfkeywords={instructions for use, elsart, document class},%
  pdfstartview=FitH,%
  bookmarks=true,%
  bookmarksopen=true,%
  breaklinks=true,%
  colorlinks=true,%
  linkcolor=blue,anchorcolor=blue,%
  citecolor=blue,filecolor=blue,%
  menucolor=blue,pagecolor=blue,%
  urlcolor=blue]{hyperref}
\else
\usepackage[%
  breaklinks=true,%
  colorlinks=true,%
  linkcolor=blue,anchorcolor=blue,%
  citecolor=blue,filecolor=blue,%
  menucolor=blue,pagecolor=blue,%
  urlcolor=blue]{hyperref}
\fi

\makeatletter
\def\elsartstyle{%
    \def\normalsize{\@setfontsize\normalsize\@xiipt{14.5}}
    \def\small{\@setfontsize\small\@xipt{13.6}}
    \let\footnotesize=\small
    \def\large{\@setfontsize\large\@xivpt{18}}
    \def\Large{\@setfontsize\Large\@xviipt{22}}
    \skip\@mpfootins = 18\p@ \@plus 2\p@
    \normalsize
}
\@ifundefined{square}{}{}
\makeatother

\newcommand{\mpr}{\textcolor{magenta}{MPR}\xspace}

\newcommand{\bfs}{\mathbf{s}}

\newcommand{\svmpr}{\textcolor{magenta}{SV-MPR}\xspace}

\begin{document}

\title{Fast gap-filling of massive data by local-equilibrium conditional simulations on GPU
}


\author{M. Lach        \and
        M. \v{Z}ukovi\v{c}  
}


\institute{Department of Theoretical Physics and Astrophysics, Institute of Physics, Faculty of Science, Pavol Jozef \v{S}af\'arik University in Ko\v{s}ice \at
              Park Angelinum 9, 041 54 Ko\v{s}ice \\
              Tel.: +421-55-2342544\\
              \email{milan.zukovic@upjs.sk}           
}

\date{Received: date / Accepted: date}

\maketitle

\begin{abstract}
The ever-growing size of modern space-time data sets, such as those collected by remote sensing, requires new techniques for their efficient and automated processing, including gap-filling of missing values. CUDA-based parallelization on GPU has become a popular way to dramatically increase computational efficiency of various approaches. Recently, we have proposed a computationally efficient and competitive, yet simple spatial prediction approach inspired from statistical physics models, called modified planar rotator (\mpr) method. Its GPU implementation allowed additional impressive computational acceleration exceeding two orders of magnitude in comparison with CPU calculations. In the current study we propose a rather general approach to modelling spatial heterogeneity in GPU-implemented spatial prediction methods for two-dimensional gridded data by introducing spatial variability to model parameters. Predictions of unknown values are obtained from non-equilibrium conditional simulations, assuming ``local'' equilibrium conditions. We demonstrate that the proposed method leads to significant improvements in both prediction performance and computational efficiency.

\end{abstract}

\section{Introduction}
\label{sec:introduction}
With the emergence and increasing frequency of massive spatio-temporal data sets, such as those collected by remote sensing technologies, scalable numerical techniques are required for their efficient processing. For example, such data often include gaps that may occur as a result of sensor malfunctions, cloud, vegetation or snow coverage, dense precipitation or other barriers separating the sensed object and the remote sensing device~\citep{Lehman04,Cole11,Bechle13,Sun17,Kadlec17}
and may have the unfavorable effect on statistical assessment of mean values and trends. Therefore, they need to be estimated to generate gapless maps of observed variables and to facilitate prompt and informed decisions~\citep{Sickles07}. 
Most of traditional interpolation methods, such as kriging~\citep{wack03}, however, are not directly applicable to such massive data due to their computational demands. Consequently, several modifications of kriging-based methods have been developed~\citep{furr06,cres08,hart08,kauf08,ingram08,zhong16,oyeb17,marco18}, in effort to increase their computational efficiency.

Very recently, a geostatistics-informed machine learning model has been suggested by~\citet{Bai21} to improve the computational performance of the ordinary kriging and inverse distance weighted regression method~\citet{emmen21} to improve the performance of the standard inverse distance weighted (IDW) method~\citep{shepard1968two}. Aiming at the same goal, a statistical physics inspired approach that employs models based on Boltzmann-Gibbs exponential joint densities has also been proposed~\citep{dth03,dthsel07,dth15,mz-dth09a,mz-dth09b,mz-dth18,hristopulos2021stochastic}. In this concept, spatial correlations are captured by means of short-range interactions, instead of the experimental variogram used in geostatistical methods, which renders the proposed interpolation methods computationally very efficient. 
Especially, the recently introduced method that employed the \emph{modified planar rotator} (\mpr) model~\citep{mz-dth18}, due to its computational efficiency (roughly linear-time computation complexity) and ability to operate autonomously without user input, was shown to be appropriate for the automated and efficient processing of massive gridded data, typical in remote sensing.

Nevertheless, spatial simulation performed in a sequential way is still computationally costly, especially in the case of simulating huge
data sets~\citep{mari10,nune10,pere15,rase15}. With new developments in hardware architecture and its availability in common PCs, in particular multi-core CPU and general purpose Graphics Processing Units (GPU), more and more popular way of overcoming the computational inefficiency is achieved by parallel implementations. 
Up to date most standard interpolation methods, including kriging and IDW, have been parallelized on high performance and distributed architectures~\citep{kerry98,cheng10,guan10,pesq11,hu15,misr20,que21,migal22} and general-purpose computing GPU~\citep{xia11,tahma12,cheng13,guti14,mei14,stoj14,mei17,marcel17,zhang18,zhan18}. It has been shown that by means of parallelization, it is possible to achieve computational acceleration achieving up to almost two orders of magnitude compared to traditional single CPU implementation. 

Parallelization of spin models simulation is achievable due to the short-range character of interactions between the spin variables. By employing a highly parallel architecture of GPUs impressive up to 1000-fold speedups can be achieved~\citep{weig11,weig12}. Our recent GPU implementation of the spin-model-based \mpr method led to almost 500-fold computational speedup, compared to single-processor calculations, for massive data sets~\citep{MPR-GPU}. Thus, using an ordinary personal computer, data sets that involve even several millions of points can be processed in a fraction of second.

Most of geostatistical methods assume spatial homogeneity/stationarity of data, even though recently a kriging-based interpolation method for non-homogeneous data has been proposed~\citep{lajaunie2020non}. However, if one targets large spatial data, in which anisotropy and non-stationarity are common, such an assumption is not justified. It is not reasonable to assume that one set of model parameters can capture scale-dependent relationships between covariates and the outcome variable that vary in space. The most common techniques for modelling such data are geographically weighted regression (GWR) and spatially-varying coefficients (SVC) methods~\citep{foth03,gelf03,harris2010use,finl11}. There are some relatively efficient GWR methods that have been developed~\citep{harr10,tran16,li19} but the scalable linear-time implementation required for application to big data sets has been proposed only very recently~\citep{mura20}. Their parallelization via the Message Passing Interface lead to a further increase in computational efficiency~\citep{li19}. Another recent approach to modeling large data with non-stationary covariance structure is based on efficient local likelihood estimation in moving windows to infer spatially varying covariance parameters~\citep{pard05,wien20}.

In the present paper we implement modifications to our previously introduced GPU-accelerated \mpr method in effort to enable modeling spatial heterogeneity/non-stationarity, essential for analyzing massive spatial data. In the GPU-implemented version this can be conveniently achieved by introducing spatial dependence to the \mpr model parameter (temperature) by the so-called double checkerboard decomposition. Then, predictions of unknown values are obtained from non-equilibrium conditional situations, assuming ``local'' equilibrium conditions corresponding to local temperatures varying in space~\citep{macg93}.

The rest of the paper is structured as follows: in Section~\ref{sec:mpr} we present an overview of the previously introduced \mpr model and its GPU implementations with both spatially-uniform and spatially-varying parameters; more details are given in~\citet{mz-dth18} and ~\citet{MPR-GPU}. The statistical and computational performance of the \mpr-based models is investigated and compared to the standard approach in Section~\ref{sec:results}. In the last Section~\ref{sec:conclusion} we summarize our findings and present conclusions.

\section{\mpr methods}
\label{sec:mpr}

\subsection{\mpr method with spatially uniform parameter}
\label{sec:su-mpr}
Let us consider a two-dimensional square grid $\mathcal{G}$ of the size $L \times L $ nodes with partially known values (samples). Let us denote locations of the samples of the spatial process $Z(\bfs)$ on the grid nodes as $\mathcal{G}_{S} = \{ \bfs_{n} \}_{n=1}^{N}$, where $N < L^2$ and their values as $\mathbf{Z}_{s} = (z_{1}, \ldots, z_{N})^\top$ (where $\top$ denotes the matrix transpose). The task is to estimate the missing values $\mathbf{\hat{Z}}_{p} = (\hat{z}_{1}, \ldots, \hat{z}_{P})^\top$
of the process at the grid nodes $\mathcal{G}_{P} = \{ \tilde{\bfs}_{p} \}_{n=1}^{P}$. Thus, the intersection of the sets $\mathcal{G}_{P}$ and $\mathcal{G}_{S}$ is empty and their union represents the full grid $\mathcal{G}$. 


In the following we briefly outline the basic idea of the \mpr method, recently introduced for efficient and automatic prediction of partially sampled non-Gaussian data on regular grids~\citep{mz-dth18}. It should be noted that, unlike geostatistical methods, the \mpr method makes no restrictive assumptions regarding the probability distribution of the spatial process. Instead, it assumes that the spatial correlations are imposed by local (nearest-neighbor) interactions between the nodes of $\mathcal{G}$.
The \mpr method employs the modified planar rotator (\mpr) spin model in the framework of a Gibbs-Markov random field (GMRF). In the first step, the original data are linearly transformed to continuously-valued ``spin'' variables (or spin angle $\phi$) space $[0, 2\pi]$. Then, conditional Monte Carlo (MC) simulations of the \mpr Hamiltonian 
\begin{equation}
\label{Hamiltonian_mod}
{\mathcal H}=-J\sum_{\langle i,j \rangle}\cos[q(\phi_i-\phi_j)],
\end{equation}
where $J>0$ is the interaction between neighboring spins and $q \leq 1/2$ is the modification parameter, are performed at the temperature $T$. The latter is estimated by matching of the specific energy of the whole grid (including sample and prediction points) with that calculated only from samples as
\begin{equation}
\label{eq:mpr-sse}
e_s = - \frac{1}{N_{SP}}\sum_{i = 1}^{N}\sum_{j \in nn(i)}\cos[q(\phi_i-\phi_j)],
\end{equation}
where $j \in nn(i)$ is the sum over the sample (non-missing) nearest neighbors of the site $\bfs_{i}$ (i.e., $\bfs_{j} \in \mathcal{G}_{S}$), and $N_{SP}$ is the total number of the nearest-neighbor sample pairs. After reaching thermodynamic equilibrium, the spin values at the prediction locations are back-transformed to the original values. Finally, spatial prediction of missing data is based on taking the mean of the respective conditional distribution at the target site given the incomplete measurements. Its high computational efficiency makes the \mpr method applicable to massive data, for example remotely sensed images. More details about the \mpr algorithm can be found in the paper by~\cite{mz-dth18}.

\subsection{\svmpr method with spatially varying parameter}
While the \mpr model has proved to be competitive for non-Gaussian data, its reliance on the single parameter - the reduced temperature $T$ - for the whole data set naturally restricts its applicability. In particular, since $T$ is related to spatial variability, a single parameter value cannot adequately capture spatial variability of the data showing some heterogeneity, a commonly present feature in massive data. For example, if the studied data set includes domains of almost constant values as well as domains with large spatial fluctuations, no single value of $T$ can be optimal for both of these regimes. The sample specific energy given by Eq.~\eqref{eq:mpr-sse} in a subsystem with nearly constant values will be very low, which will result in $T$ taking values close to $0$. Such values would not be representative for a subsystem with high variability and vice-versa. The sample specific energy calculated from all data (including domains with low and high variability) and subsequently estimated temperature thus may not be characteristic for either of the involved domains. As a result, the spatial variability of the predictions in the domains with low (high) variability will be over- (under)-estimated. 

\subsubsection{GPU-implemented \svmpr method}
Generally, such a problem can be addressed by introducing a spatial dependence into the model parameters, as it is done in the GWR approach~\citep{foth03,gelf03}. The current implementation in CUDA offers a very straightforward way to do this for the \mpr method. Parallel computations in CUDA are facilitated by virtual processors called threads that can be mapped to hardware resources to perform computations for individual (or multiple) data points or spins. These threads are organized into blocks consisting of up to 1024 threads arranged in one, two or three dimensions (here, we use square blocks). Thread blocks are in turn organized into a grid of up to three dimensions. Block and grid dimensions are specified at the launch of each kernel - a function which executes computations on the GPU. More information about the CUDA programming model and the implementation of the \mpr algorithm on the GPU is given in our previous paper \citep{MPR-GPU}. 

\subsubsection{Block-specific parameter inference and simulation}
The energy of the system is calculated using the parallel reduction algorithm~\citep{parallel_reductions, faster_parallel_reductions}. Each thread is mapped to a single spin and computes the energy of a bond with two of its neighbors using the Hamiltonian (\ref{Hamiltonian_mod}) in such a way that all the bonds present in the system are counted. Then, the results of each thread within a given block are \textit{reduced} (summed) to a single value representing the energy contribution of that block. Since at this point, the energy is known for each block, same energy matching procedure as described in Sect.~\ref{sec:su-mpr} can be used to assign an individual value of $T$ to each block. Thus, we obtain multiple values of the simulation parameter $T$ for different regions of the system. In the extreme case, when a large percentage of the block's data is missing and there are no bonds between samples (no nearest neighbors) for energy calculation and temperature estimation, such a block will be assigned the temperature corresponding to the median value of the available block temperatures.

Since during simulation, threads representing individual spins are also grouped into blocks, instead of a single global value of $T$, each thread can use the value which corresponds to the block it belongs to. One must make sure that the same configuration of thread blocks used for the energy computation is also used for the Metropolis update, otherwise spins from one part of a system may end up using a temperature based on a different part of the system with diverse behavior. This approach synergizes with an optimization technique called double checkerboard (DC) decomposition \citep{weig12}, which is an extension of the single checkerboard (SC) decomposition described in our previous paper \citep{MPR-GPU}.

\begin{figure}[t!]
\centering
\includegraphics[scale=0.4,clip]{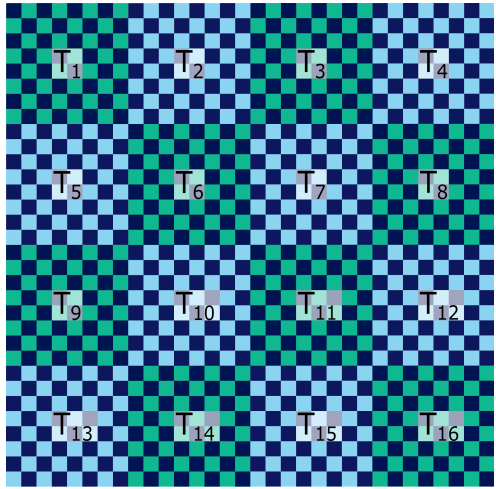}
\caption{Schematic representation of a double checkerboard decomposition of the spin grid with block specific temperatures $T_i$, $i=1,\hdots,16$. Each small square represents a grid node with its associated spin variable. Each thread performs calculations for a single spin. Sub-grid A comprises the dark nodes, while sub-grid B includes the light nodes. The grid is decomposed further into larger dark and light tiles corresponding to individual thread blocks responsible for the numerical operations on the spins included within the tile. Each tile can be simulated using a different temperature computed only from the spins within the tile.}
\label{fig:DC}
\end{figure}

Assuming the data are localized on a two dimensional square grid (generalization to any regular grid is straightforward), they can be split into two sets sitting on two interpenetrating sub-grids, e.g., A and B. The nearest neighbors of any node on the sub-grid A belong to the sub-grid B, and vice versa. Therefore, the updating algorithm can be applied to all the spins on the same sub-grid in parallel. In Fig. \ref{fig:DC}, the two sub-grids are depicted using light and dark \textit{small} squares. Our computation kernels call one thread per each sub-grid spin.

The DC decomposition splits the grid further into larger tiles, which correspond to thread blocks. At each MC step, first only the even (odd) numbered tiles are updated in parallel and then only the odd (even) tiles. Each tile can be loaded into the block's shared memory, which is orders of magnitude faster compared to the GPU's global memory, and we can perform multiple Metropolis updates in quick succession only within the even numbered blocks, before doing the same for the odd numbered blocks. As mentioned above, since the temperature is computed individually for each block, we can assign a different value to each tile of the decomposed grid, as depicted in Fig. \ref{fig:DC} using light and dark \textit{large} squares. In our case, this is the main reason for using the DC decomposition as we do not apply the update procedure multiple times for individual tiles. Nevertheless, due to the fact that the tiles are first loaded into shared memory, looking up the neighboring spins uses shared instead of global memory and we still get a small performance benefit, as will be discussed below. More information about the various types of memory available on GPUs can be found in our previous paper \citep{MPR-GPU} and in the CUDA programming manual \citep{nvidia_cuda_2021_guide}. We call this version of the algorithm using the DC decomposition as the \svmpr method with block-specific temperatures (BST).

\subsubsection{Site-specific parameter inference and simulation}
\begin{figure}[t!]
\centering
\includegraphics[scale=0.4,clip]{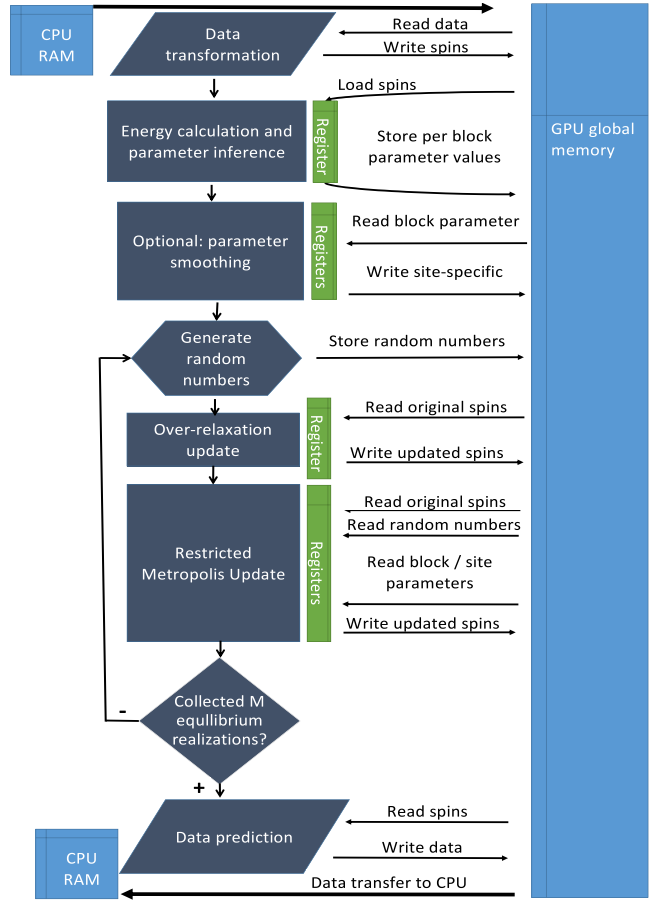}
\caption{Flowchart of the main computational steps and memory transactions performed on the GPU.}
\label{fig:flowchart}
\end{figure}

Our second approach is motivated by the effort to eliminate the undesirable block boundary effect that occurs in the above \svmpr method with BST. It results from the fact the temperature varies on the grid discontinuously as a step function and thus neighboring blocks assigned markedly different temperature values will be characterized with markedly different spatial variations of the simulated values, which can generate unnatural edges between blocks on the prediction map. Such an effect can be partially reduced by decreasing the size of the blocks. However, this approach has some limitations as very small blocks may lack samples either completely or their reduced numbers within blocks may lead to imprecise block-specific temperature estimation. An alternative approach allows us to reduce the block size so that each of them contains only one spin and thus to obtain a smooth variation of temperatures on the grid. Assigning a value of $T$ to each spin individually can be viewed as a limiting case of decreasing the block size $l_b \rightarrow 1$. However, instead of actually decreasing the block size we apply a simple smoothing algorithm to the block temperatures. The latter can be achieved by starting from the BST state and recursively replacing the temperature at each site with an average value of the surrounding area with some radius $r_s$. To reach the desired level of smoothness, it can be applied $n_s$ times in succession. We will refer to this version of the algorithm as the \svmpr method with site-specific temperatures (SST). The implementation of the \svmpr algorithms on the GPU is illustrated in the flowchart shown in Fig.~\ref{fig:flowchart}.

\section{Results}
\label{sec:results}
\subsection{Data}

\begin{figure}[t!]
\centering
\subfigure{\includegraphics[scale=0.3,clip]{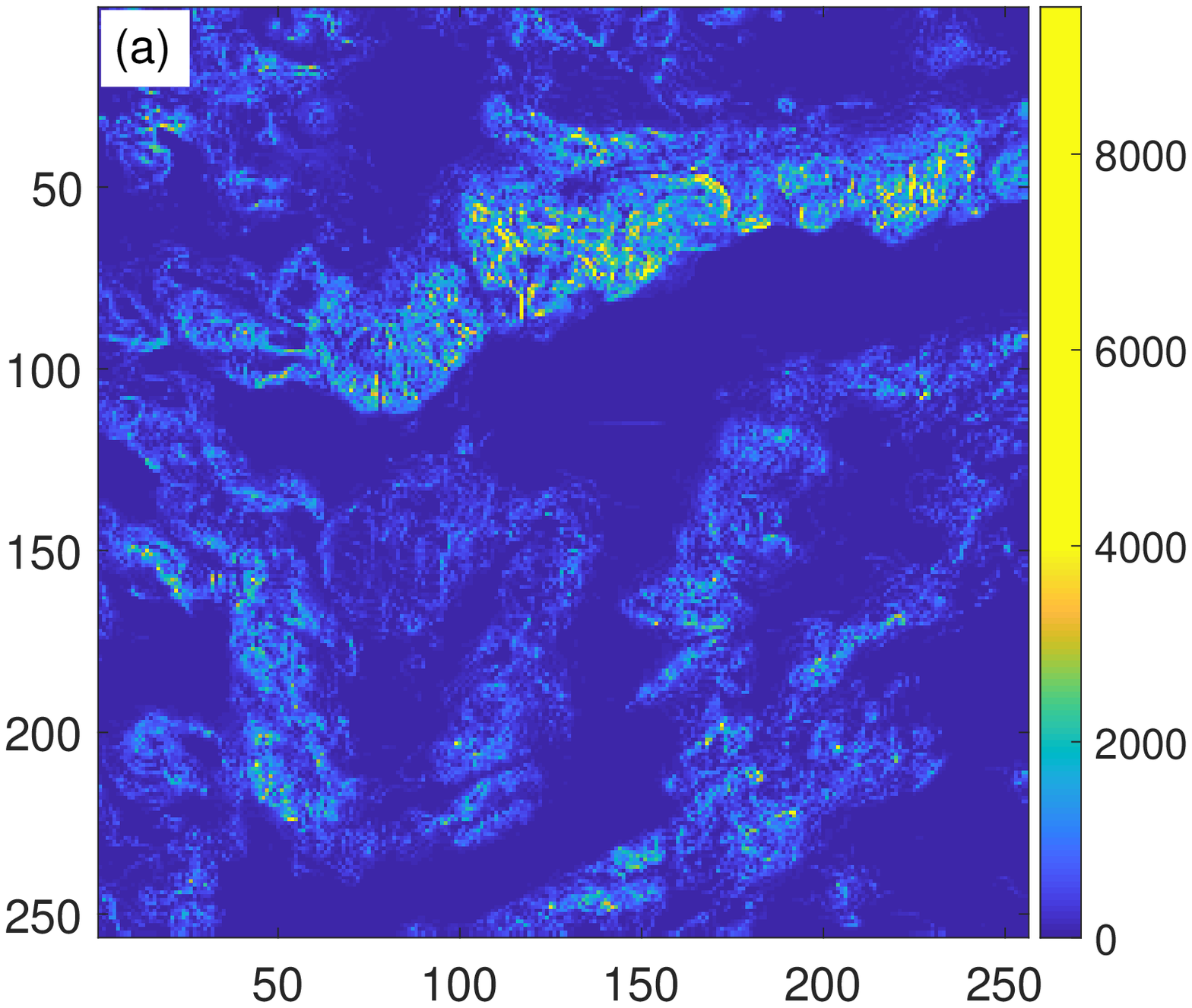}\label{fig:orig_walker_lake}}
\subfigure{\includegraphics[scale=0.3,clip]{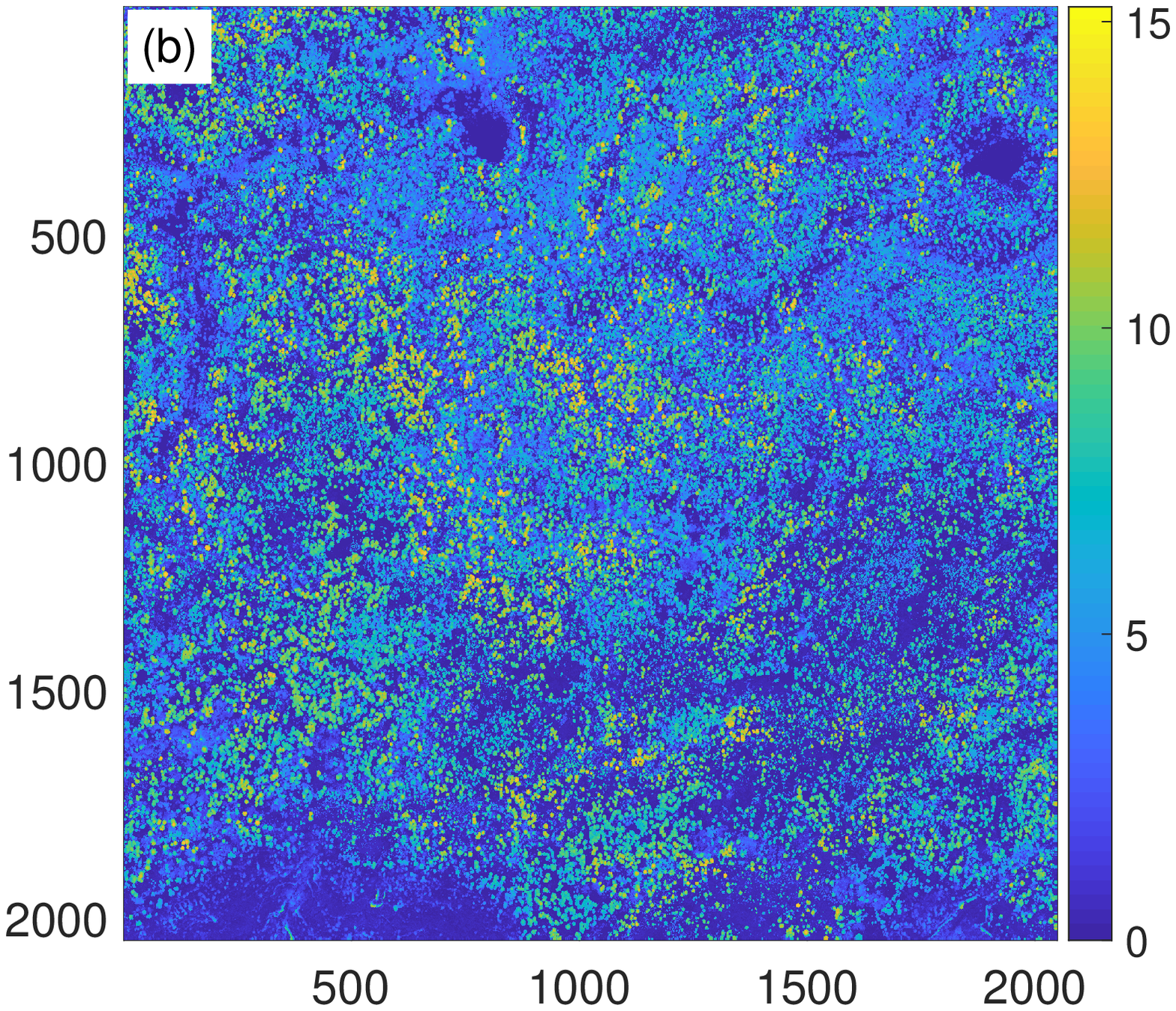}\label{fig:orig_kaibab_plateau}}
\subfigure{\includegraphics[scale=0.3,clip]{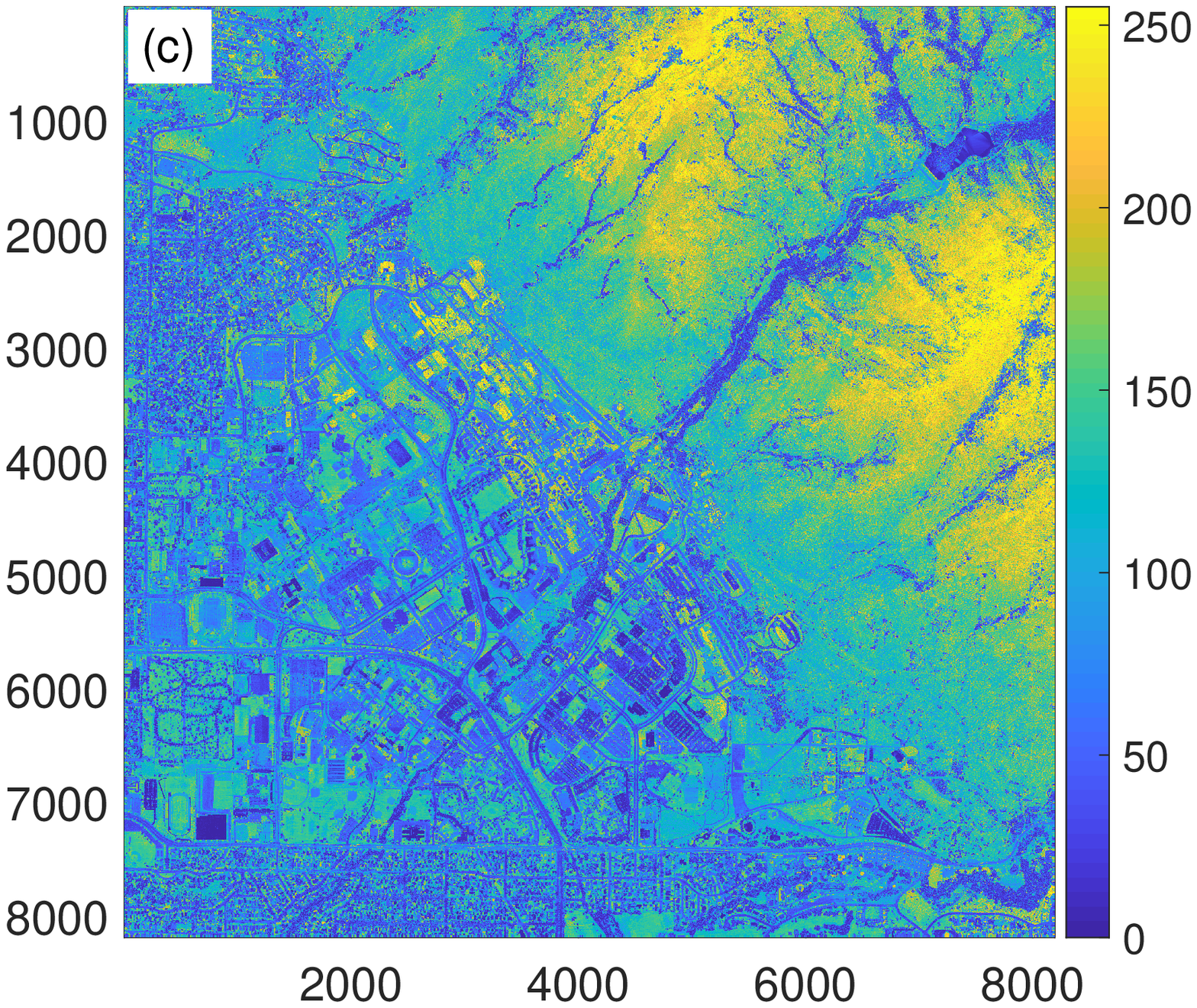}\label{fig:orig_wasatch_front}}\\
\caption{Spatial distributions of (a) Walker lake, (b) Kaibab plateau and (c) Wasatch front.
}
  \label{fig:colormaps}
\end{figure}

\begin{figure}[t!]
\centering
\subfigure{\includegraphics[scale=0.37,clip]{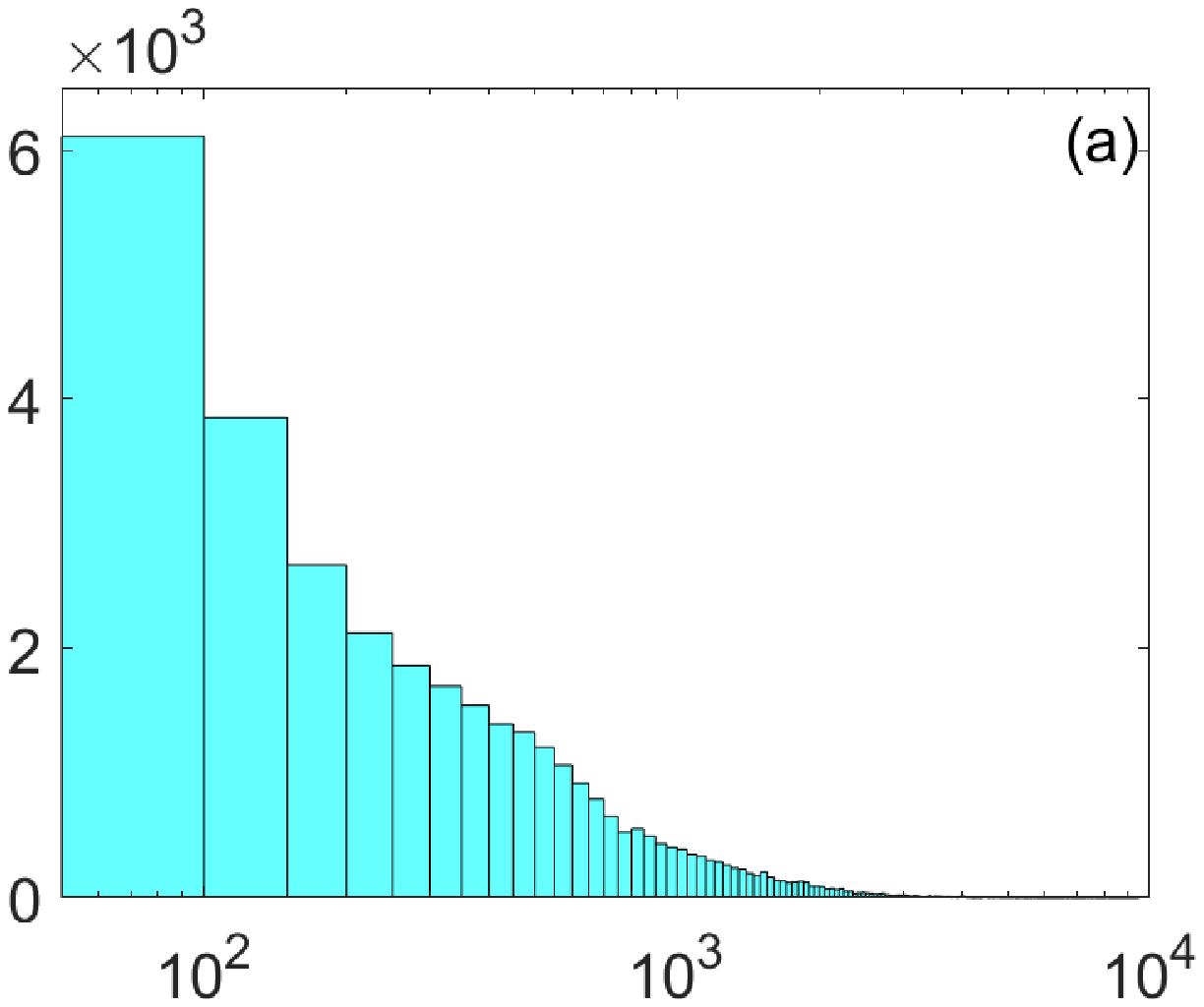}\label{fig:hist_walker_lake}}
\subfigure{\includegraphics[scale=0.37,clip]{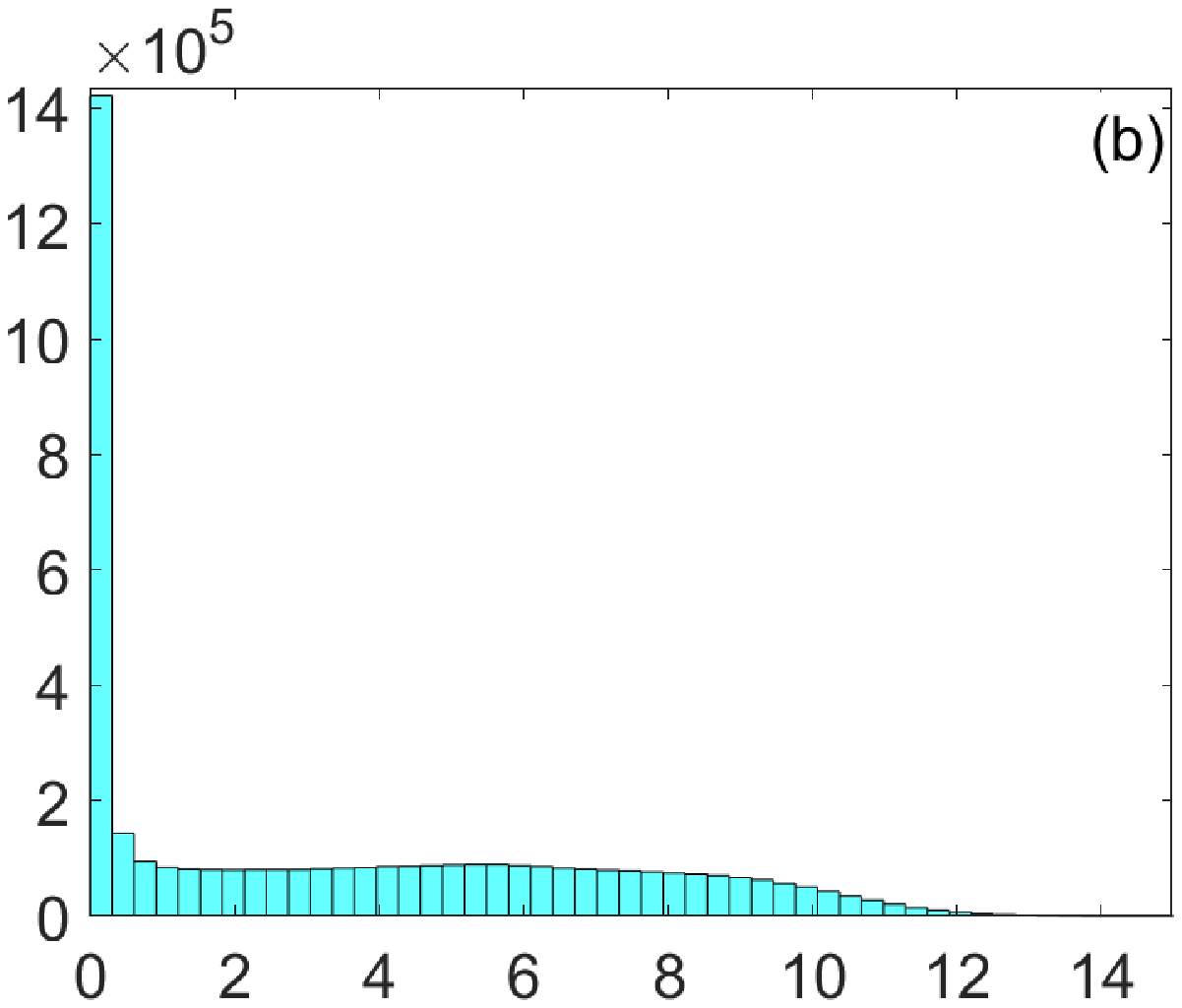}\label{fig:hist_kaibab_plateau}}
\subfigure{\includegraphics[scale=0.37,clip]{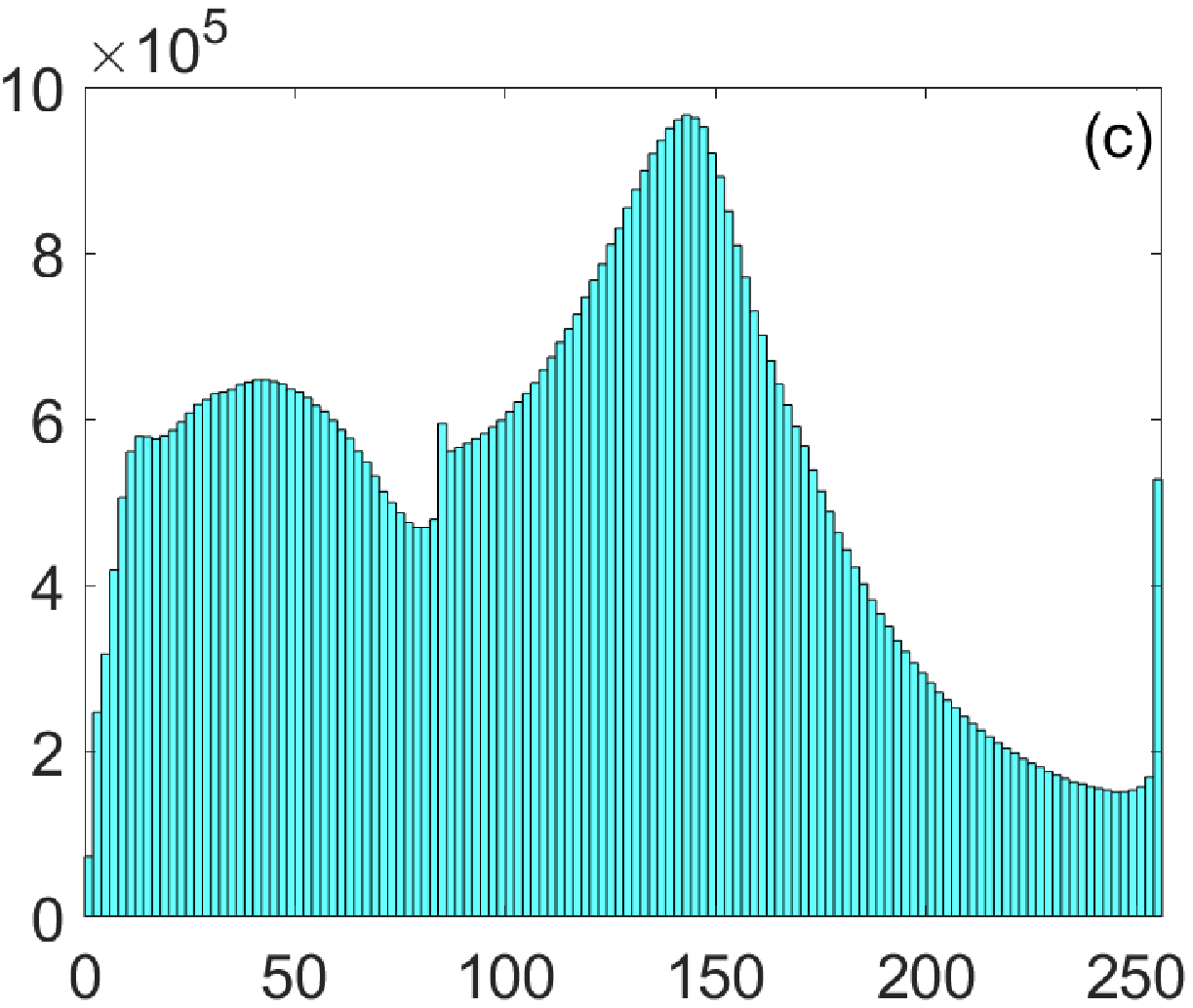}\label{fig:hist_wasatch_front}}\\
\caption{Histograms of (a) Walker lake, (b) Kaibab plateau and (c) Wasatch front.
}
  \label{fig:histograms}
\end{figure}

To assess the effect of introducing the spatial dependence into the model parameter, we compare both the prediction performance and computational efficiency of the present \svmpr model with the original \mpr model. The comparison is performed on several big real world data sets showing a heterogeneous character of their spatial variability (see Fig.~\ref{fig:colormaps}) and non-Gaussian (skewed, multimodal, etc.) distributions (see Fig.~\ref{fig:histograms}). The first one represents the synthetic pollutant concentration data derived from a digital elevation model of the Walker lake area in Nevada~\citep{Applied-geostatistics}. A map showing a 2D projection of the pollution field is presented in Fig.~\ref{fig:orig_walker_lake} along with the histogram on a semi-log scale in Fig.~\ref{fig:hist_walker_lake}. The units used for the $Z$ values are arbitrarily set to parts per million (ppm). The map shows the presence of both larger dark regions with the values close to zero and almost no variability, as well as brighter regions with very large values and relatively high variability\footnote{Note the adjusted color map, showing all the values $Z \gtrsim 4\ 000$ in yellow, to better visualize the extreme data in the tail.}. The histogram shows that the distribution is highly positively skewed indicating the predominance of very low concentrations with just a small portion of extremely large values.

Further, we considered much larger data sets with the linear grid sizes $L = 2\ 048$ and $L = 8\ 192$, collected using airborne light detection and ranging (LIDAR) technology. They represent the canopy height (in meters) of the forests around Kaibab plateau, Arizona \citep{kaibab_plateau_dataset} (Figs.~\ref{fig:orig_kaibab_plateau} and \ref{fig:hist_kaibab_plateau}), and the digital surface model of Wasatch front, Utah~\citep{wasatch_front_dataset} (Figs.~\ref{fig:orig_wasatch_front} and~\ref{fig:hist_wasatch_front}). These data sets include values over extensive spatial domains with nontrivial distributions, as shown in the histograms in Fig.~\ref{fig:histograms}. 
The statistical properties of all the data sets are summarized in Table~\ref{table:stats}.

\begin{table}[h!]
\caption{Summary statistics of the used data sets.}
\label{table:stats}
\begin{tabular}{lccccccc}
\hline\noalign{\smallskip}
\hfil Dataset & \hfil Number of points & \hfil Range & \hfil $\bar{z}$ & \hfil $z_{0.50}$ & \hfil $\sigma_z$ & \hfil skewness & \hfil kurtosis \\
 \hline
  Walker lake & $65\,535$ ($L=256$) & $ 0-9\ 500$&    $ 289$&   $ 64 $&   $ 516$&    $ 3.6 $&    $ 23.3 $\\
  Kaibab plateau & $4\,194\,304$ ($L=2\ 048$) & $ 0-15$&    $ 3.6$&   $ 2.7 $&   $ 3.5$&    $ 0.58 $&    $ 2 $\\
  Wasatch front  & $67\,108\,864$ ($L=8\ 192$) & $ 0-255$&    $ 112$&   $ 117 $&   $ 62$&    $ 0.13 $&    $ 2.2 $\\
 \noalign{\smallskip}\hline
\end{tabular}
\end{table}

\subsection{Prediction validation}
To evaluate the performance of the \mpr-based prediction algorithms, we simulate missing values by setting aside a portion of the complete data to be used as a validation set. We typically generate $M = 100$ different configurations by randomly removing between $p = 30\% - 90\%$ (or $p=0.3-0.9$) of data points. The \mpr predictions are based on the conditional mean as evaluated from the conditional MC simulation. The reconstructions are compared with the true values, first by visually inspecting the reconstructed data and then statistically, using two validation measures: the average absolute error (AAE) defined as
\begin{equation}
{\rm AAE} = \frac{1}{P}  \sum_{\textbf{r}_p \in G_p} |\epsilon(\textbf{r}_p)|,
\label{eq:AAE}
\end{equation}
and the root average squared error (RASE)
\begin{equation}
{\rm RASE} = \sqrt{\frac{1}{P}  \sum_{\textbf{r}_p \in G_p} \epsilon^2(\textbf{r}_p)},
\end{equation}
where $\epsilon(r_p) = Z(\textbf{r}_p) - \hat{Z}(\textbf{r}_p)$ is the difference between the true value $Z(\textbf{r}_p)$ and the predicted value $\hat{Z}(\textbf{r}_p)$ at the site $\textbf{r}_p$ and $P = pL^2$ is the number of prediction sites. Both of these quantities are then averaged over the $M=100$ different sample configurations to calculate the mean AAE (MAAE) and mean RASE (MRASE). CUDA-based calculations are executed on a PC with NVIDIA GeForce RTX 2080 SUPER GPU, using CUDA version 10. The CPU host system is equipped with 8-core 3 GHz Intel(R) Core(TM) i7-9700F CPU with  32 GB RAM, running Ubuntu 20.04.3 LTS Linux.

\subsection{Visual inspection of reconstructions}
\subsubsection{Standard \mpr method}
\begin{figure}[t!]
\centering
\subfigure{\includegraphics[scale=0.42,clip]{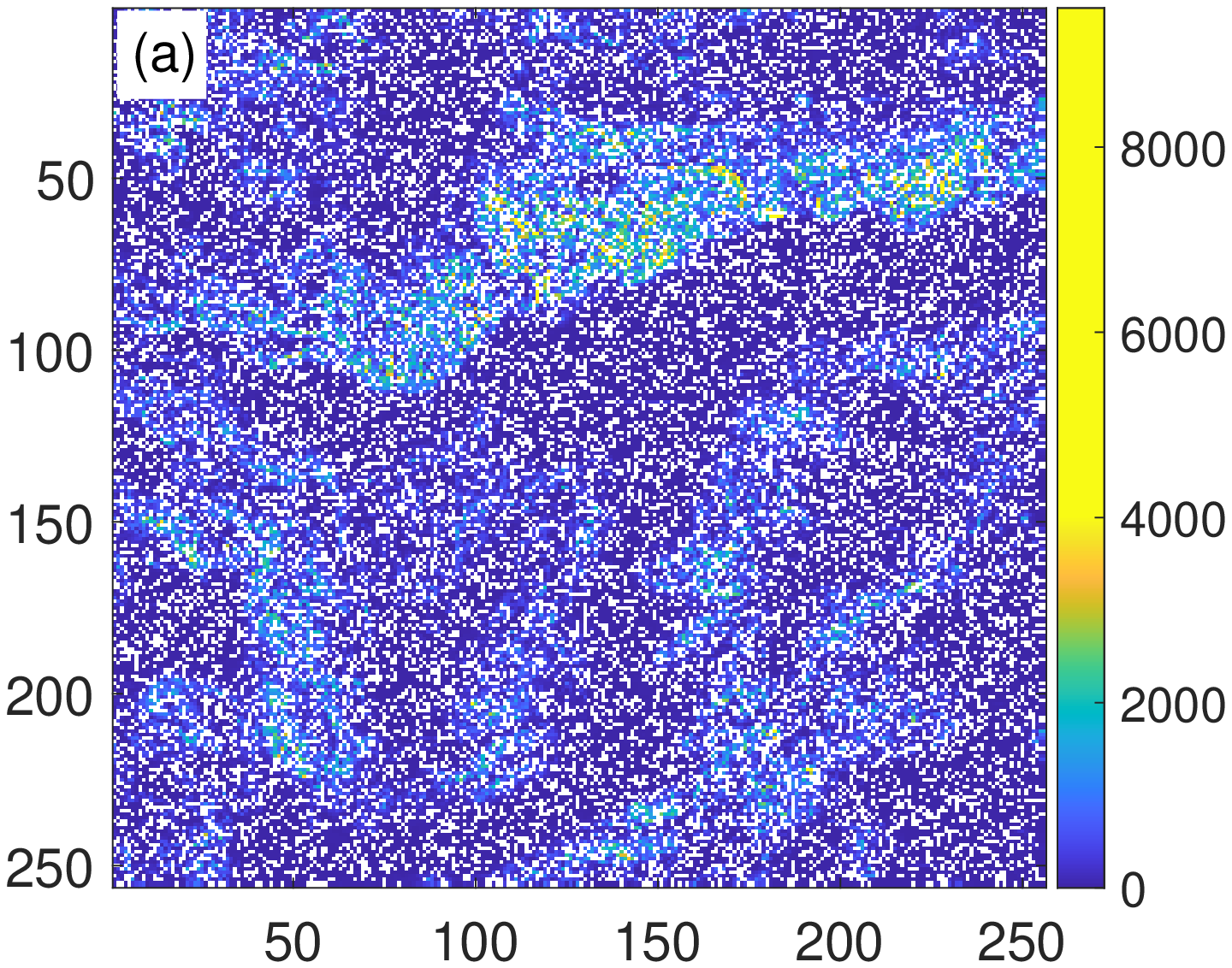}\label{fig:missing_data_030}}
\subfigure{\includegraphics[scale=0.42,clip]{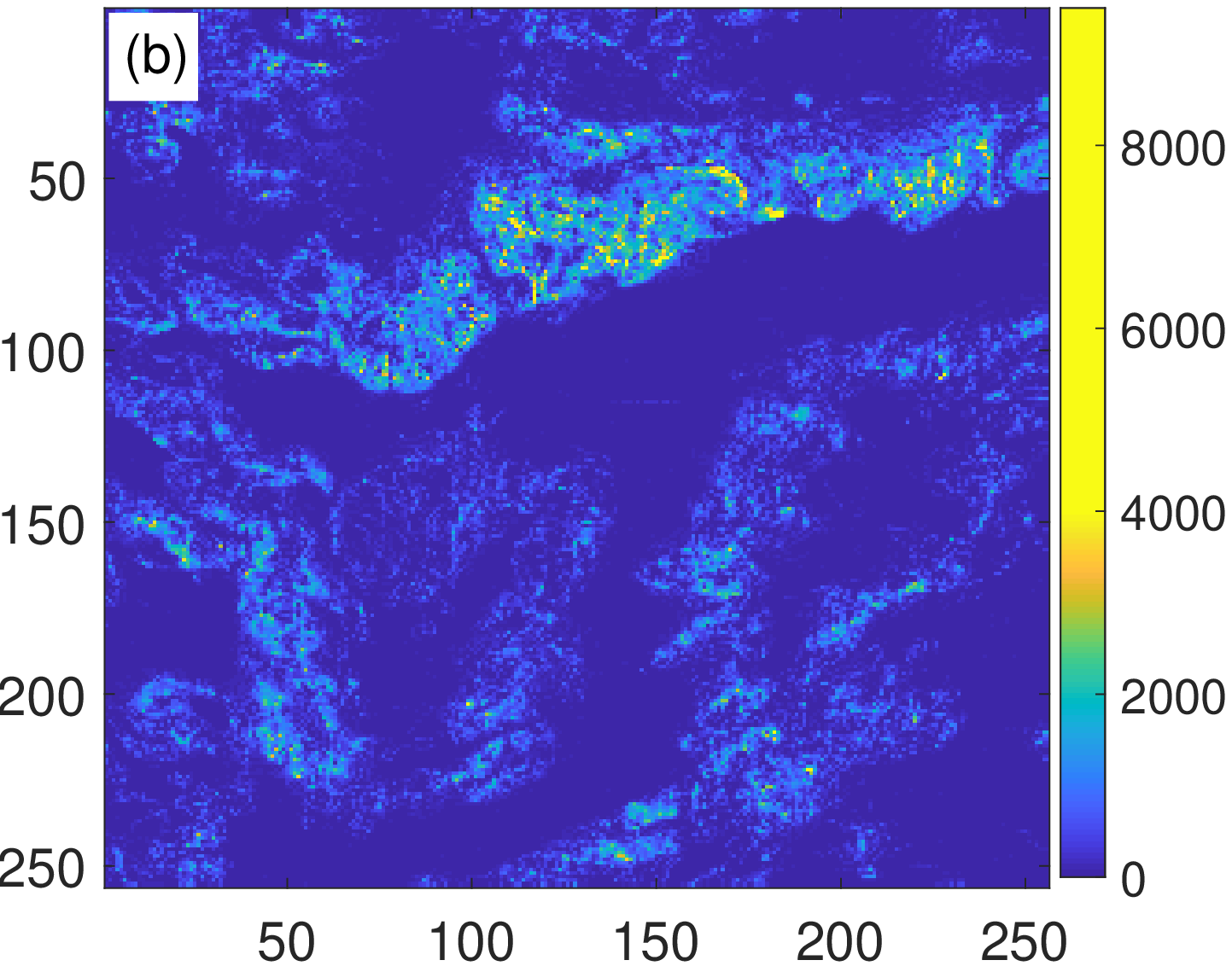}\label{fig:recons_SC_030}}\\
\subfigure{\includegraphics[scale=0.42,clip]{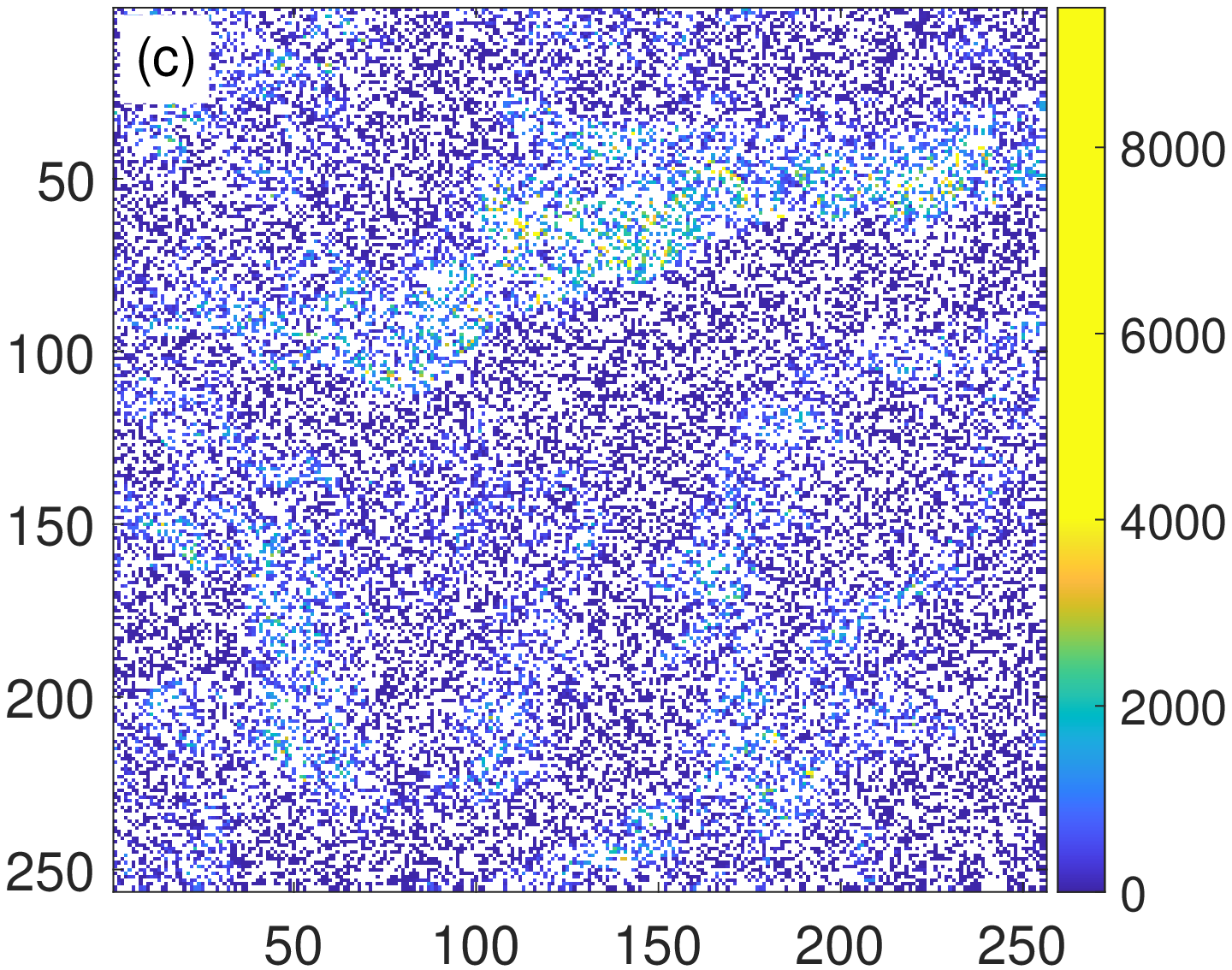}\label{fig:missing_data_060}}
\subfigure{\includegraphics[scale=0.42,clip]{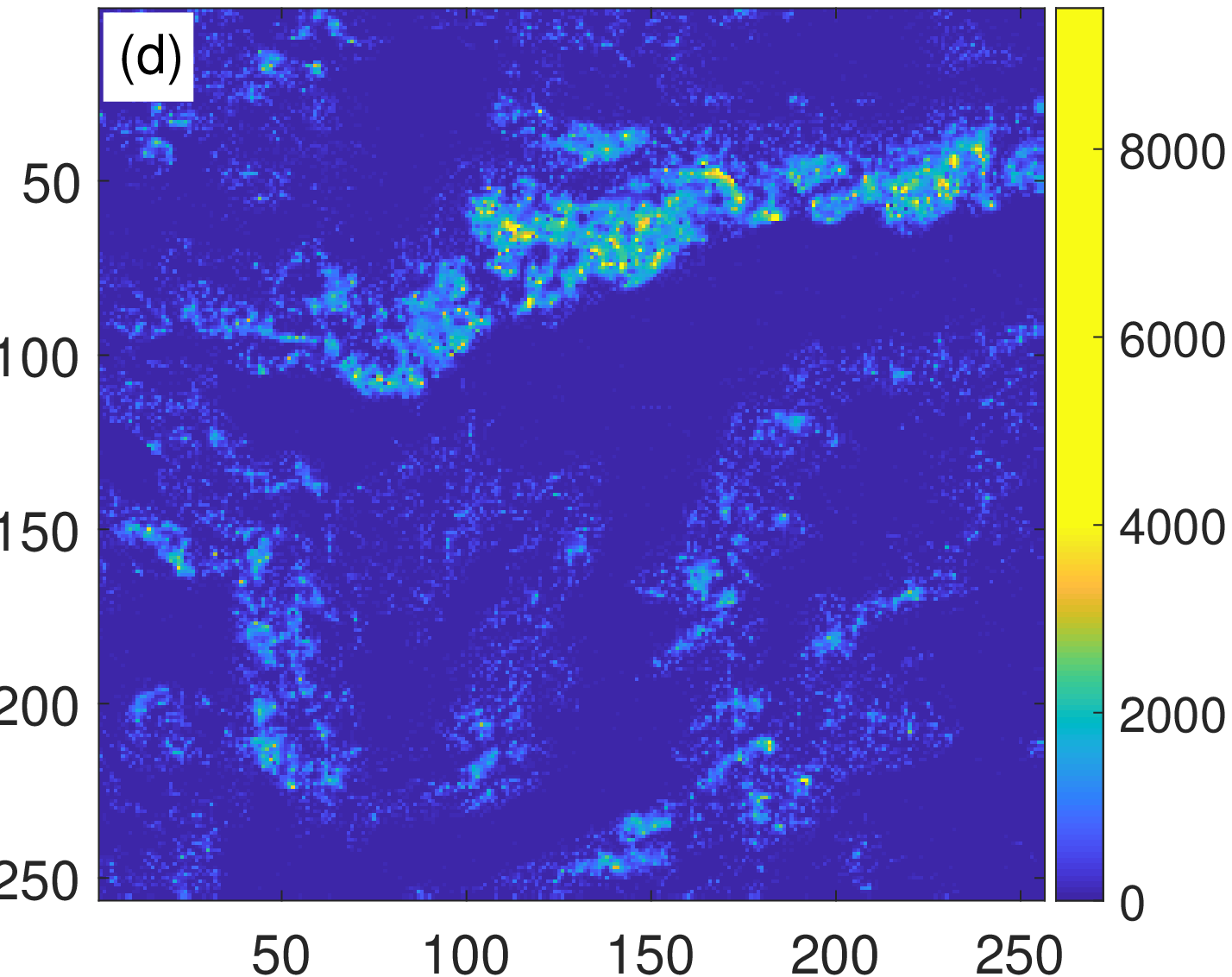}\label{fig:recons_SC_060}}\\
\subfigure{\includegraphics[scale=0.42,clip]{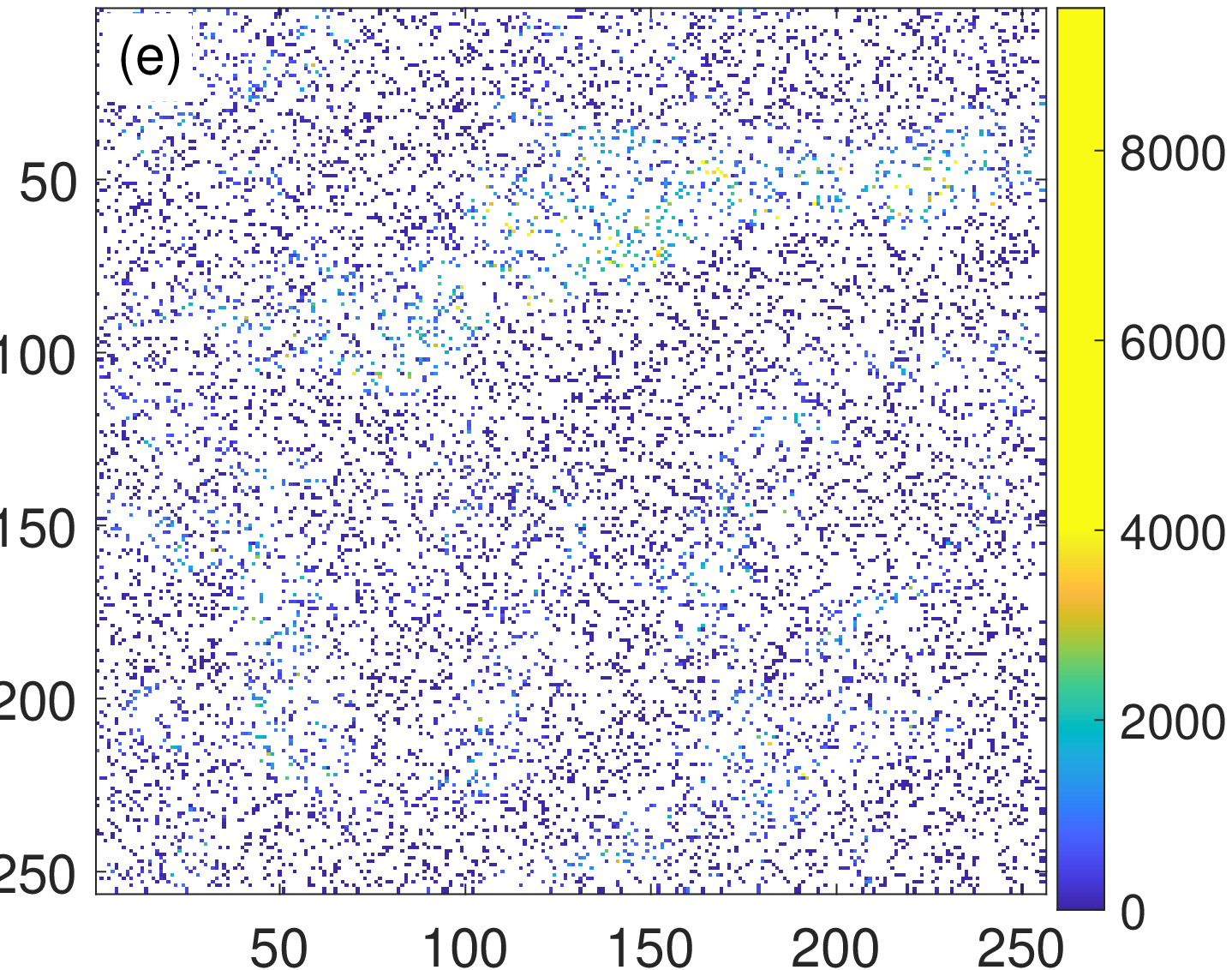}\label{fig:missing_data_085}}
\subfigure{\includegraphics[scale=0.42,clip]{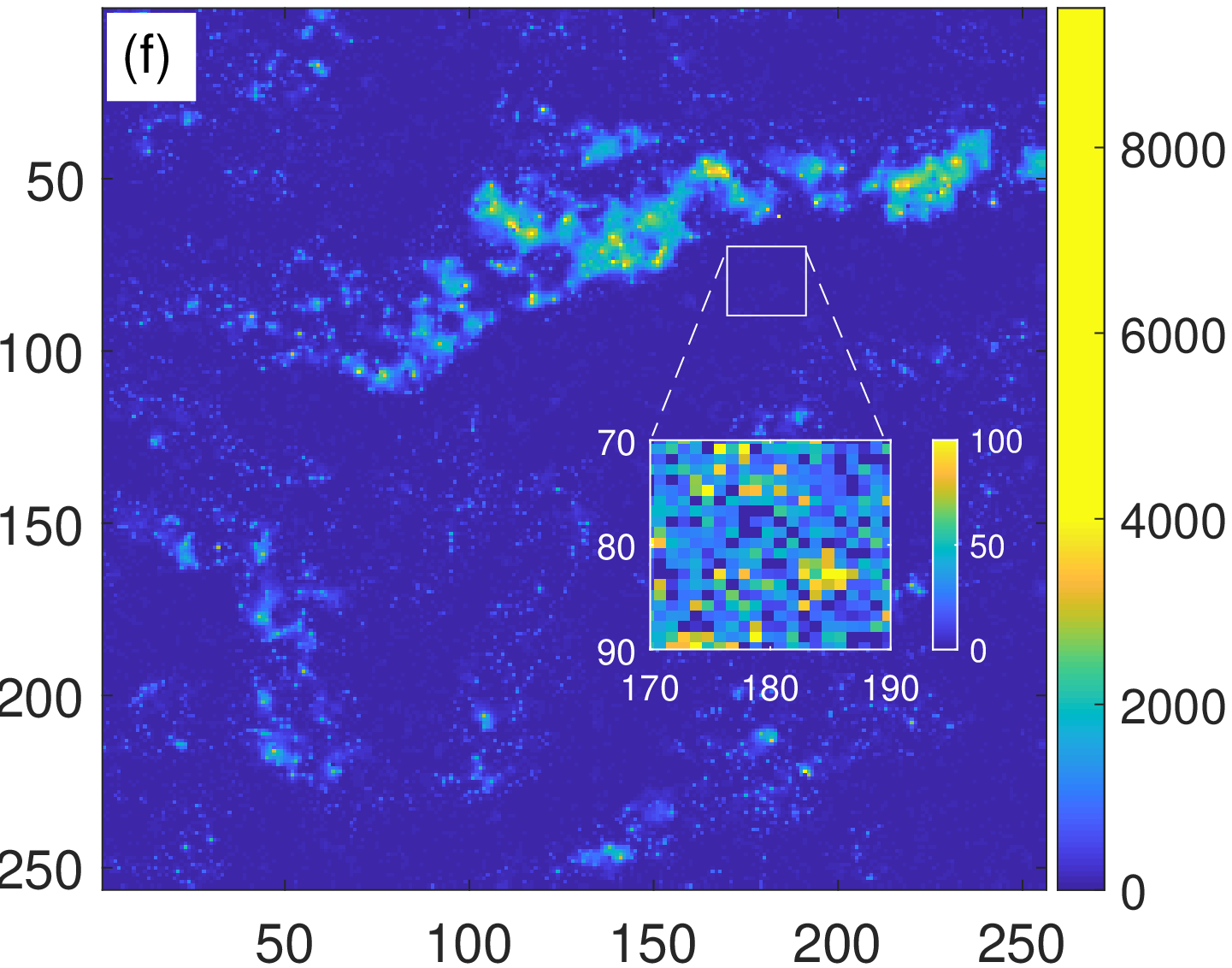}\label{fig:recons_SC_085}}\\
\caption{Original Walker lake data with (a) $p=30\%$, (c) $p=60\%$ and (e) $p=85\%$ of the data (white color) randomly removed and the corresponding \mpr-based reconstructions obtained at the (spatially uniform) temperatures (b) $\hat{T}=0.0378$, (d) $\hat{T}=0.0378$ and (f) $\hat{T}=0.0672$, respectively.}
\label{fig:SC}
\end{figure}
Let us first demonstrate the performance of the original \mpr algorithm with a single (spatially uniform) parameter $T$ and its weak point, especially when applied to the data with highly heterogeneous spatial variability, such as the Walker lake data set. The gappy data along with their reconstructions using the \mpr algorithm are depicted in Fig.~\ref{fig:SC} for three values of the thinning ratio $p=30\%,60\%$ and $85\%$. The \mpr algorithm performs fairly well for small values of $p$, thanks to the abundance of the conditioning sample data. However, with the increasing sparsity of the samples the prediction performance quickly deteriorates. In particular, the predicted values in the regions corresponding to very low or zero concentrations (dark regions) noticeably overestimate the true values, which is reflected in the appearance of speckles with lighter colors. On the other hand, the extremely large values are greatly underestimated, albeit it is less conspicuous in the reconstructed map due to the scarcity of such data. This phenomenon becomes much more pronounced at larger concentrations of missing data (lack of conditioning sample data), such as for $p=85\%$ presented in Fig.~\ref{fig:recons_SC_085}. This problem is not conspicuous on the large scale but it is evident if we zoom in a small area, as shown in the inset. In this particular area the true values are $Z(\textbf{r}_p)=0$, nevertheless, the predictions $\hat{Z}(\textbf{r}_p) \in [0,100]$. Considering the fact that the \mpr model includes only one parameter, this averaging effect is not surprising. The temperature $\hat{T}$ is estimated based on all the samples, involving regions with different degrees of spatial variability and, thus the resulting mean value of $\hat{T}$ cannot be representative in all these areas and thus cannot reflect the local conditions. 

\subsubsection{\svmpr method - BST implementation}
\begin{figure}[t!]
\centering
\subfigure{\includegraphics[scale=0.42,clip]{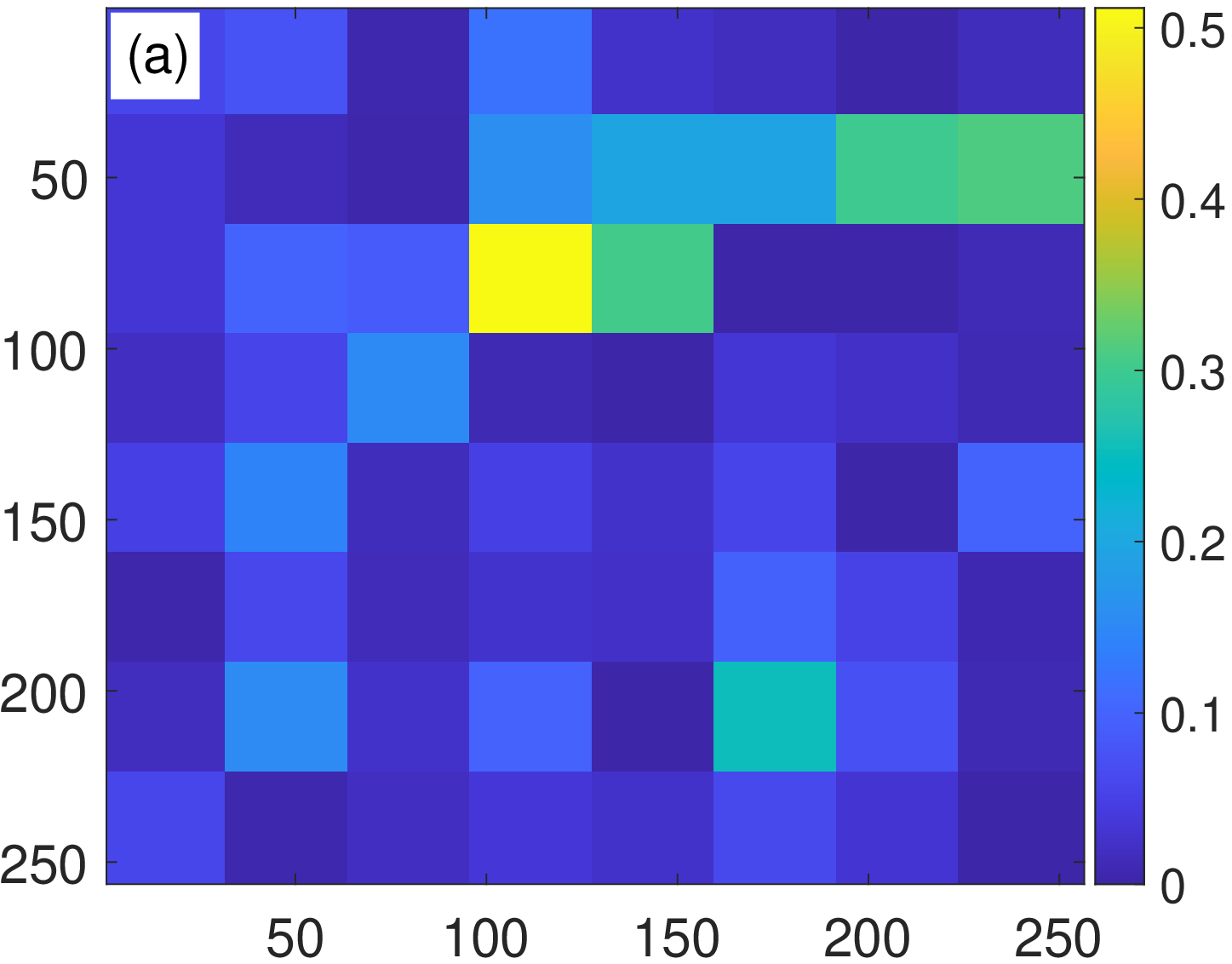}\label{fig:block_32_temps}}
\subfigure{\includegraphics[scale=0.42,clip]{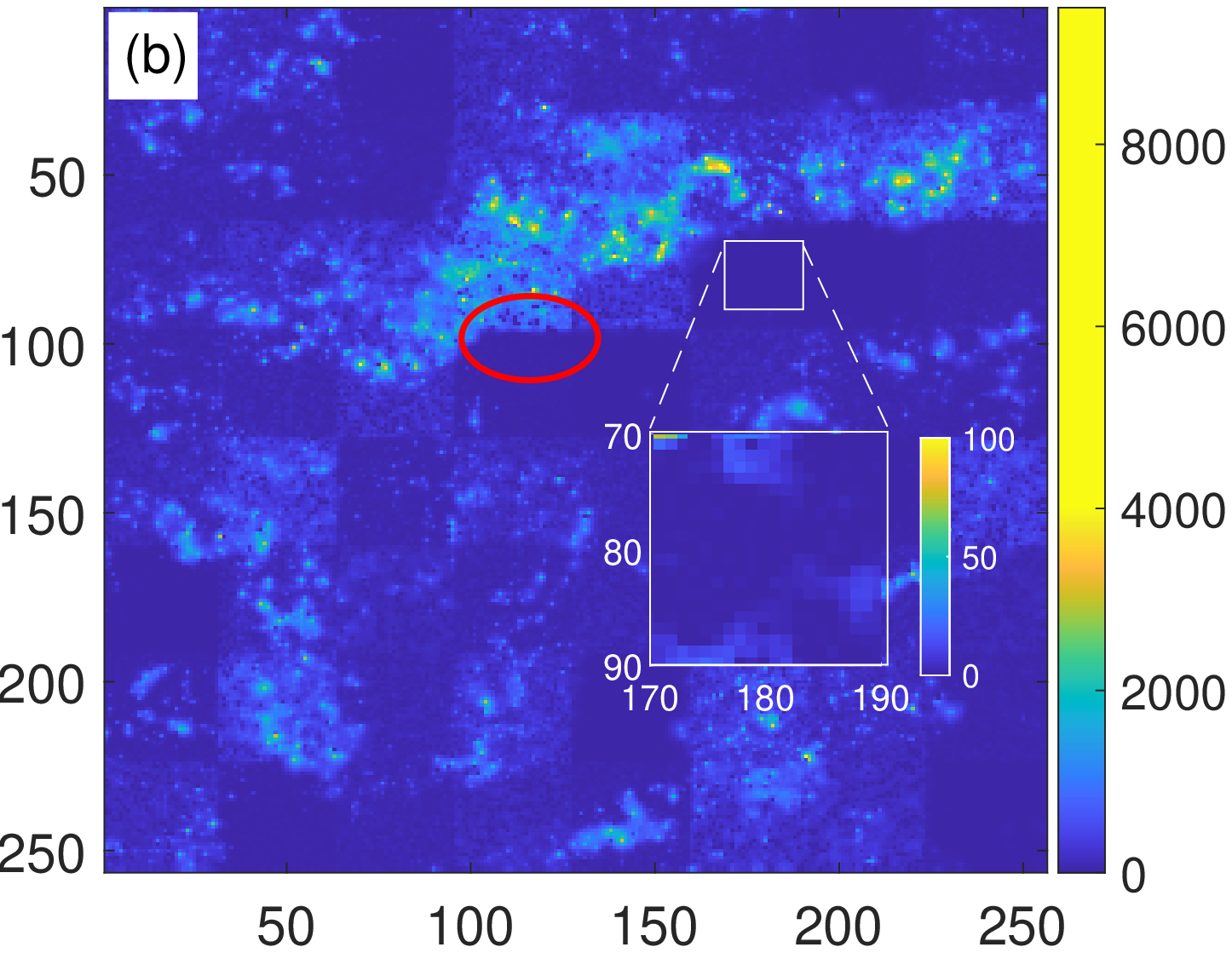}\label{fig:recons_DC32}}\\
\caption{(a) Distribution of block-specific temperatures and (b) map of reconstructed data using the BST implementation of the \svmpr method for $l_{b} = 32$. The red ellipse encloses an example of the edge-like artifact of using block-specific temperatures and the square area demonstrates the improved \svmpr predictions compared to the \mpr ones in the same area shown in Fig.~\ref{fig:recons_SC_085}.}
  \label{fig:DC32}
\end{figure}
To best demonstrate the beneficial effects of implementing spatial variability of the parameter in the presently introduced \svmpr versions of the algorithm, we chose the high sample sparsity case of $p = 0.85\ (85\%)$. The \mpr reconstruction is depicted in Fig.~\ref{fig:recons_SC_085} and the corresponding (spatially uniform) parameter value, estimated from the specific energy matching principle based on all samples, is $\hat{T}=0.0672$. By using the double checkerboard decomposition in the \svmpr implementation the spatial distribution of the estimated block-specific temperatures (BST) with square blocks of the linear size $l_b = 32$ is shown in Fig.~\ref{fig:block_32_temps}. One can witness a great variability in BST, the values of which correlate with the sample variation in the respective blocks. In particular, the blocks with almost constant sample values close to zero are assigned very low values of $\hat{T} \approx 0$, while those with spatially highly variable samples are assigned much higher values of up to $\hat{T} \approx 0.5$. It is worth noticing that the mean value of $\bar{\hat{T}}=0.0693$, calculated over all blocks, coincides rather well with the \mpr estimate $\hat{T}$. Consequently, one can expect that the spatially-variable parameter can better model the local data variability than the spatially-uniform one. Indeed, performing simulations using the \svmpr (BST) implementation with the block specific temperatures yields the reconstruction, which suffers much less from the averaging effect than the \mpr method, as demonstrated in Fig.~\ref{fig:recons_DC32}. Visually, the reconstruction is much closer to the original data than that of the original \mpr algorithm, especially in the (dark) regions with low spatial variability. For comparison, the inset shows the \svmpr predictions in the same area as for the \mpr method in the inset of Fig.~\ref{fig:recons_SC_085}, for which $\hat{Z}(\textbf{r}_p) \in [0,20]$. On the other hand, undesirable artifacts resulting from the presence of boundaries between blocks with different parameters appear (see, e.g., the area marked by the red ellipse in Fig.~\ref{fig:recons_DC32}). Such unnatural edge-like effects are likely to emerge between blocks which include sample data with distinct degrees of variability. We note that the edges are partially smeared due to the fact that the spins at the shared boundaries of the neighboring blocks interact with each other and thus propagate fluctuations from their blocks to the surrounding blocks. Nevertheless, such a diffusion has a local character with rather limited range and cannot eliminate the edge-like effects completely.
\begin{figure}[t!]
\centering
\subfigure{\includegraphics[scale=0.42,clip]{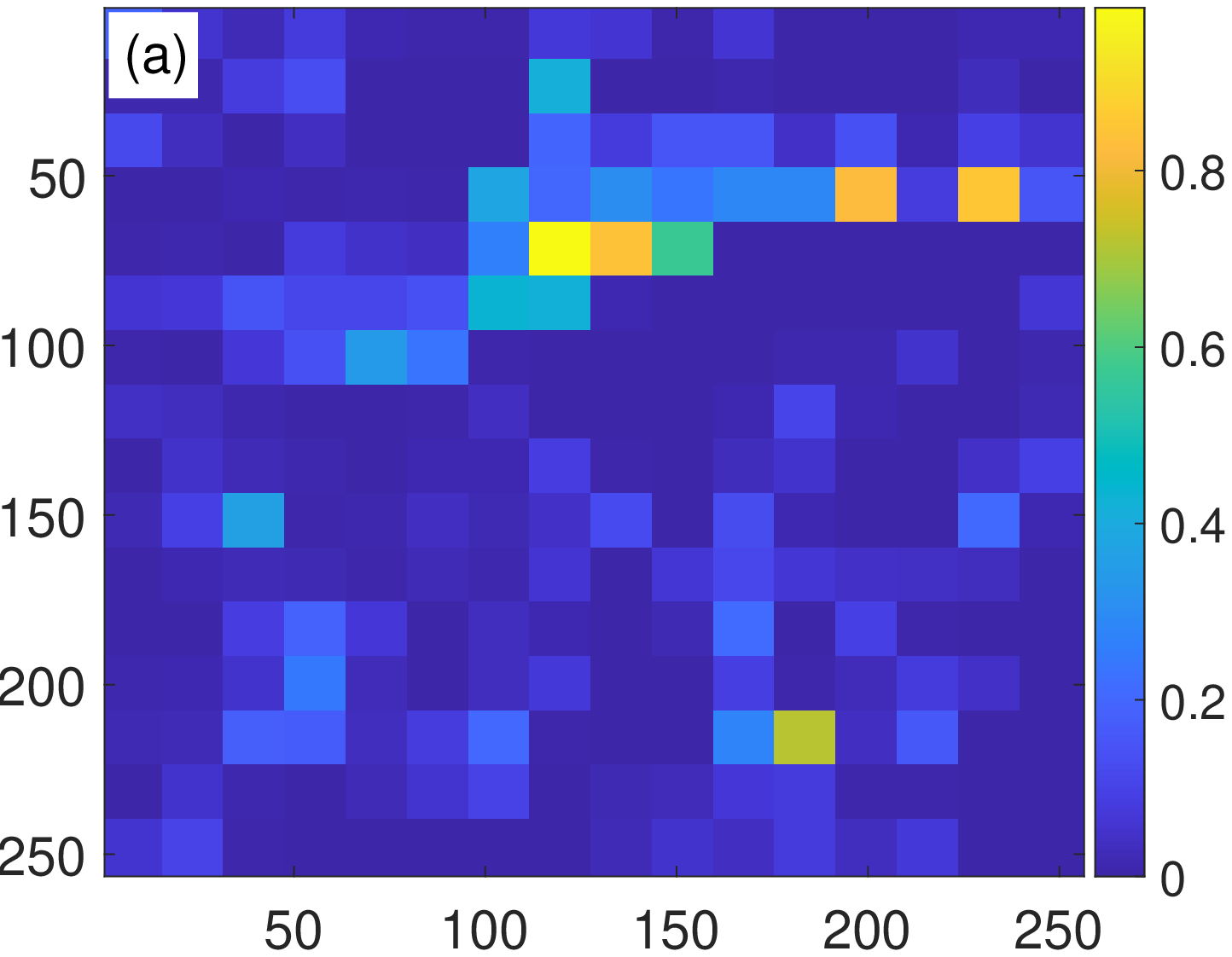}\label{fig:block_16_temps}}
\subfigure{\includegraphics[scale=0.42,clip]{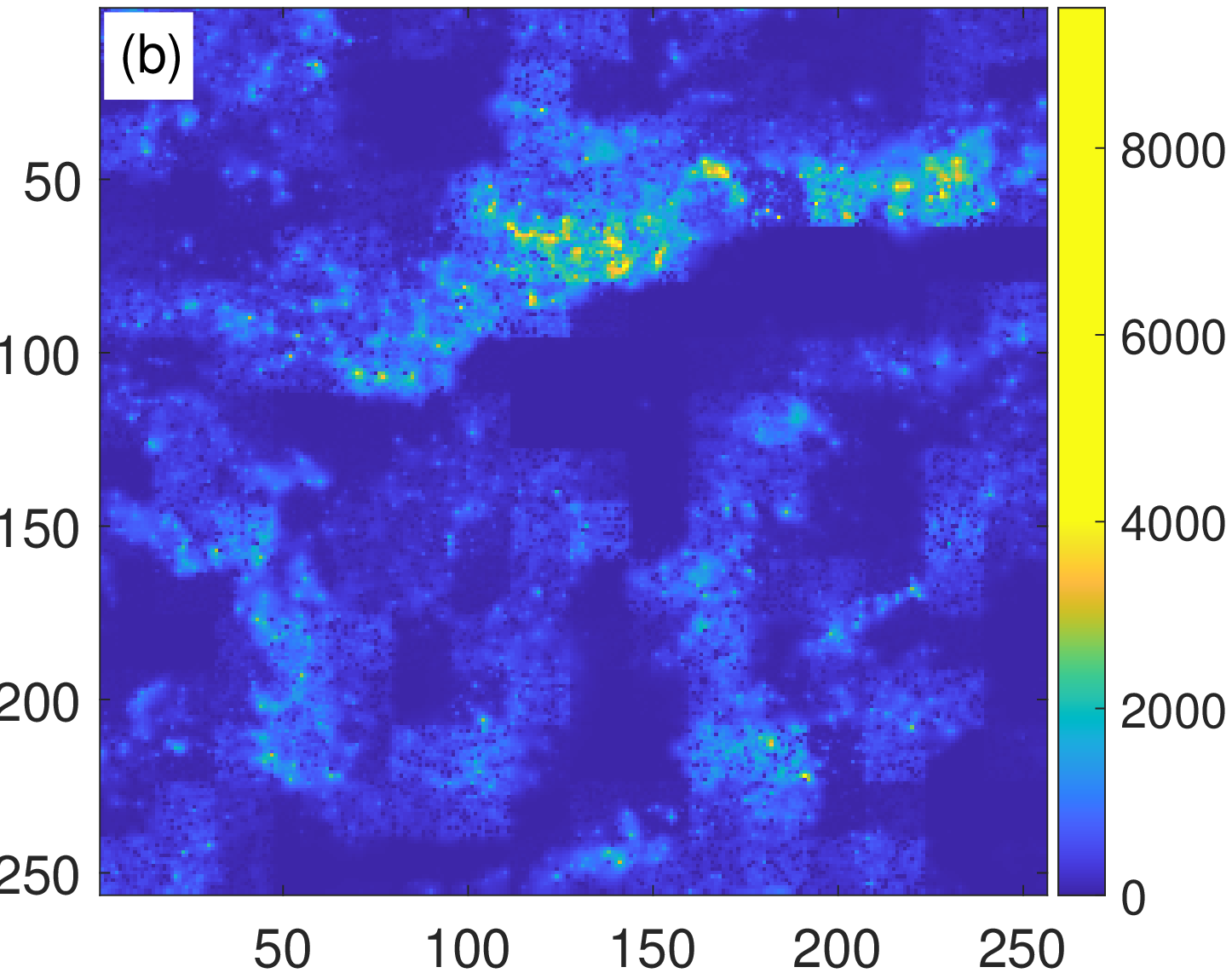}\label{fig:recons_DC16}}\\
\subfigure{\includegraphics[scale=0.42,clip]{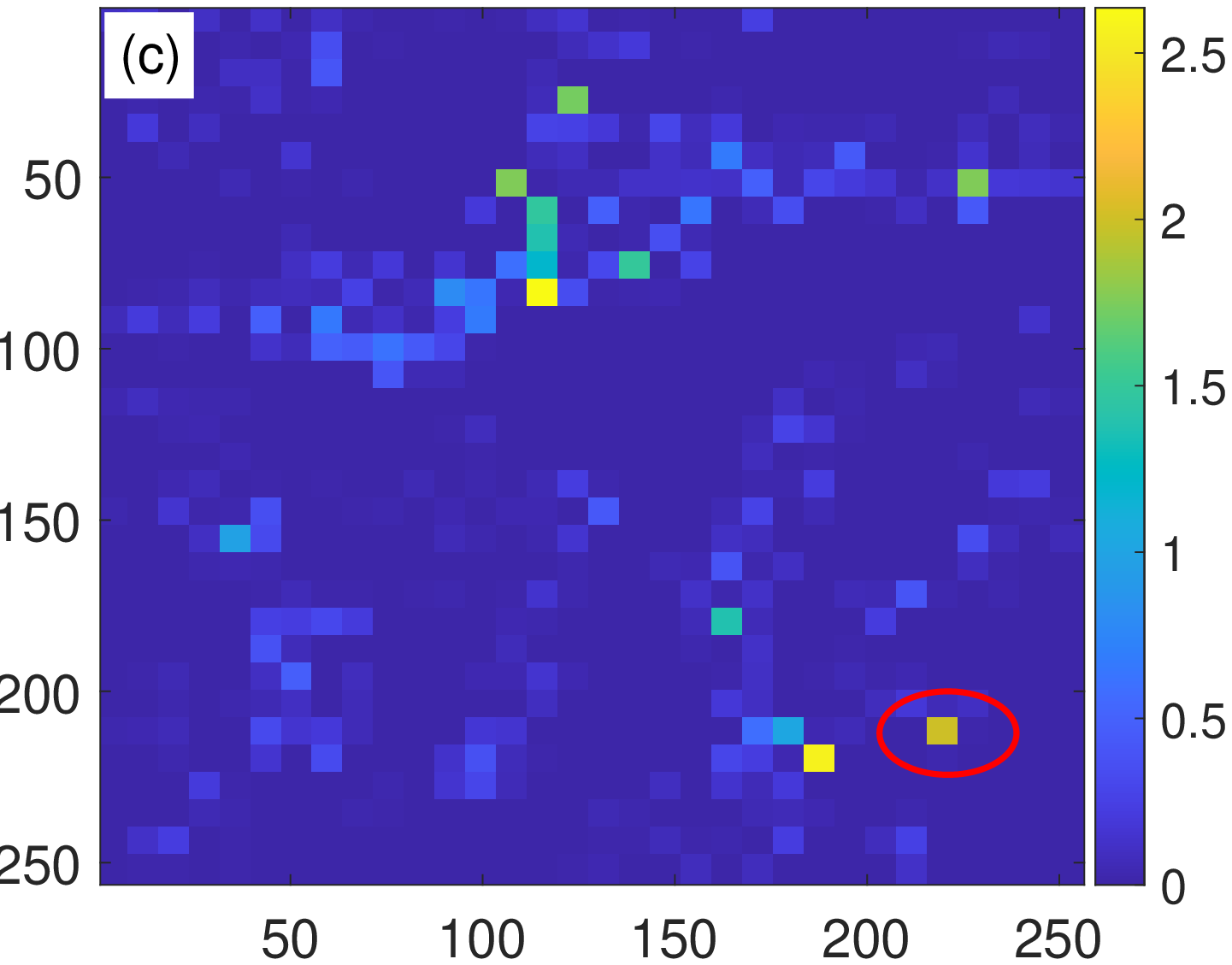}\label{fig:block_8_temps}}
\subfigure{\includegraphics[scale=0.42,clip]{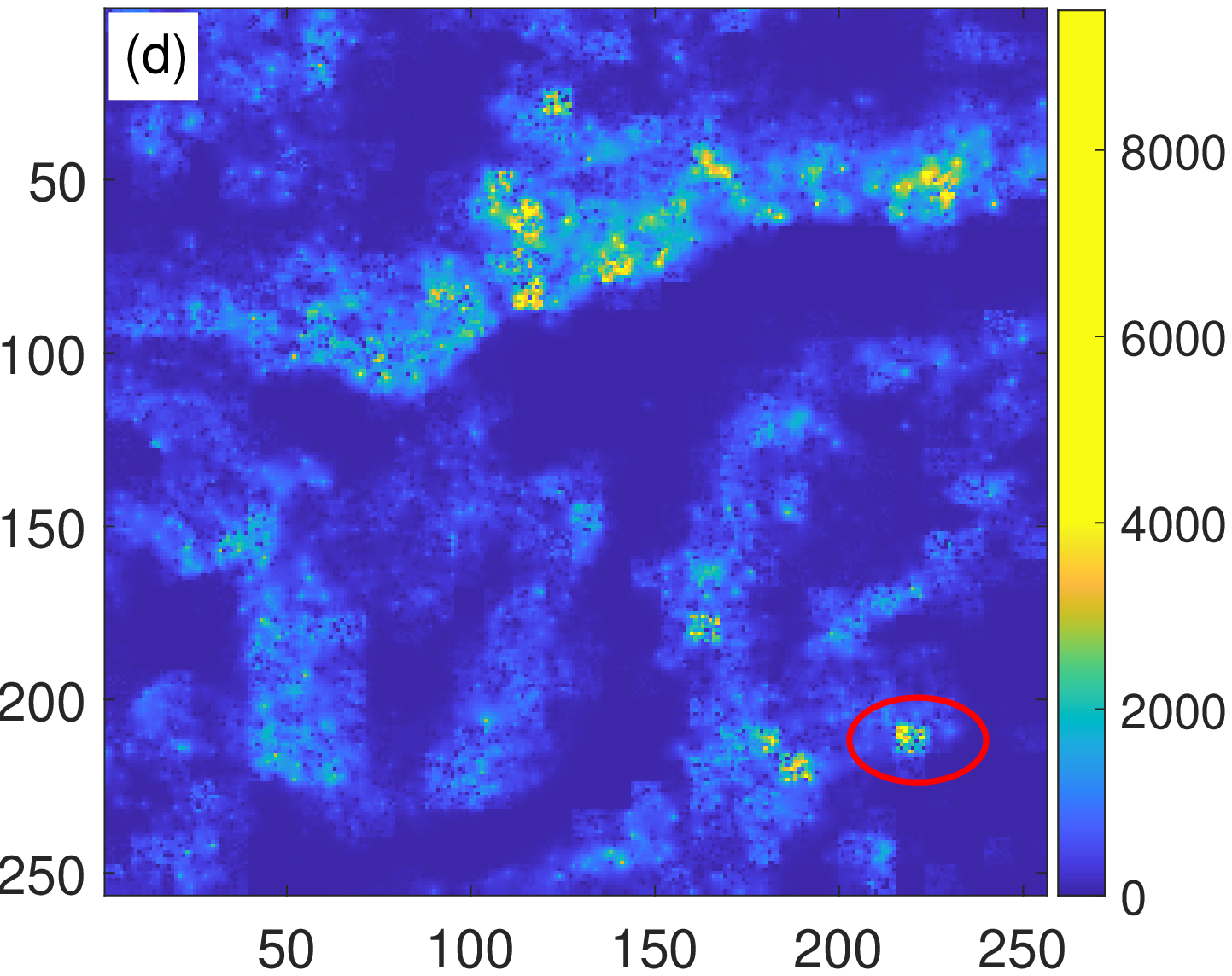}\label{fig:recons_DC8}}\\
\caption{Distributions of block-specific temperatures (left column) and the corresponding data reconstructions (right column) using the \svmpr (BST) implementation for $l_b = 16$ (upper row) and $l_b = 8$ (bottom row). The red ellipses in the lower panels highlight the area of misestimated parameter (panel (c)) and consequently data variability (panel (d)) due to lack of samples in the corresponding block.}
\label{fig:recons_DC16_8}
\end{figure}
To further eliminate these undesirable artifacts, we tried two approaches. The first one consists in gradually decreasing the linear block size $l_b$ from 32 to 16 and 8. The block temperatures (left column) and the corresponding data reconstructions (right column) are shown in Fig.~\ref{fig:recons_DC16_8}, for $l_b = 16$ (top row) and $l_b = 8$ (bottom row). Decreasing the block size leads to partial elimination of the conspicuous edges between blocks and also allows a greater flexibility in capturing the local variability. On the other hand, the decreasing block size also reduces the amount of the neighboring sample pairs (bonds) available for the calculation of the block-specific sample energies, and thus inhibits a reliable estimation of the block-specific temperatures. Particularly for very sparse samples with high values of $p$, this may cause insufficient statistics for a reliable estimation of the block-specific temperatures. This leads to the second type of the artifacts in the form of misestimation of the data variability due to the erroneous parameter estimation (see, e.g., the bright square in the lower right corner of Fig.~\ref{fig:recons_DC8} showing unexpectedly large variability including extremely large values due to the overestimation of $\hat{T}$). Notwithstanding, the presented results also demonstrate that decreasing granularity of the blocks results in a smoother spatial variation of the parameter and considerably suppresses the block boundary effects.
\subsubsection{\svmpr method - SST implementation}

\begin{figure}[t!]
\centering
\subfigure{\includegraphics[scale=0.42,clip]{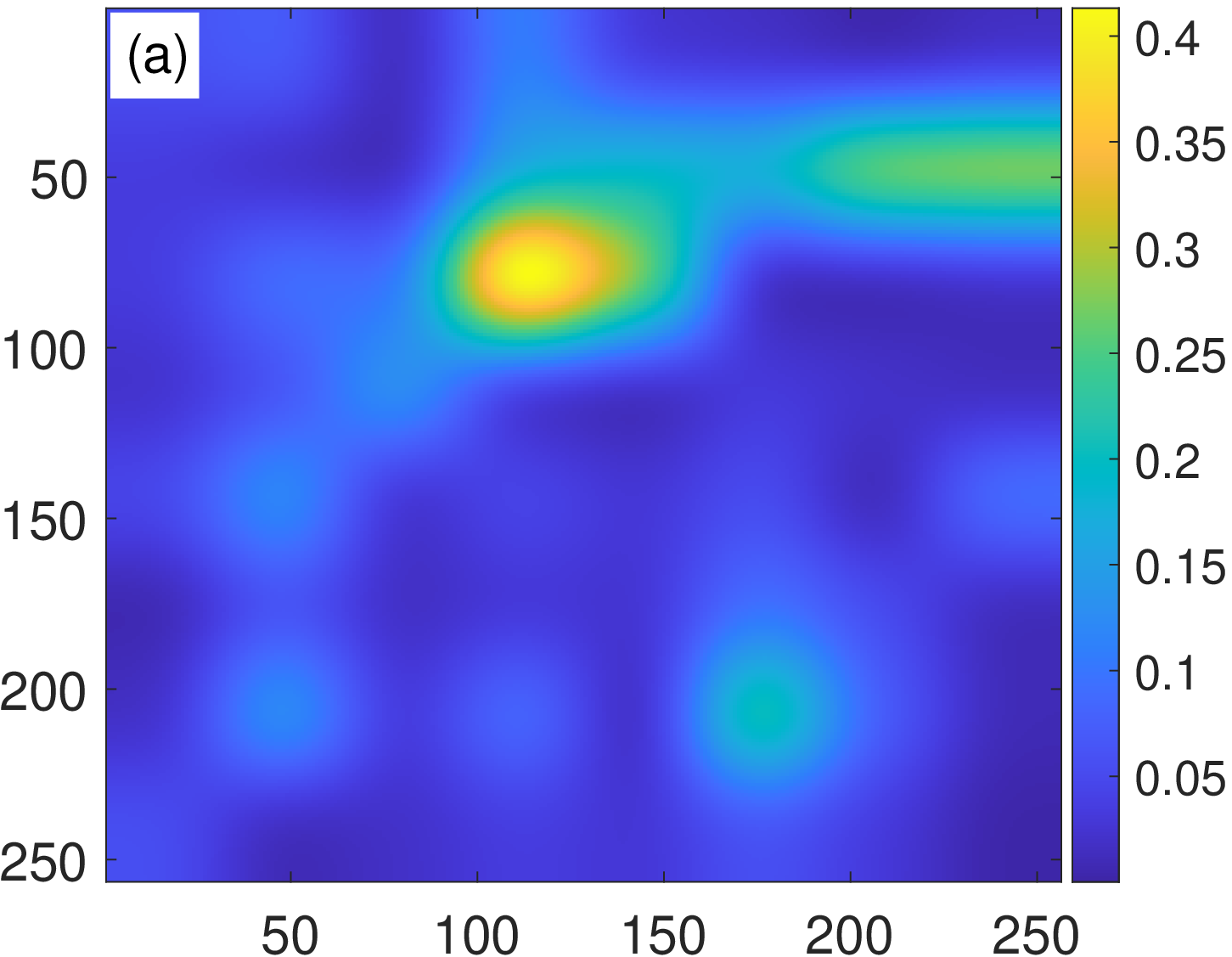}\label{fig:block_32_smooth_temps}}
\subfigure{\includegraphics[scale=0.42,clip]{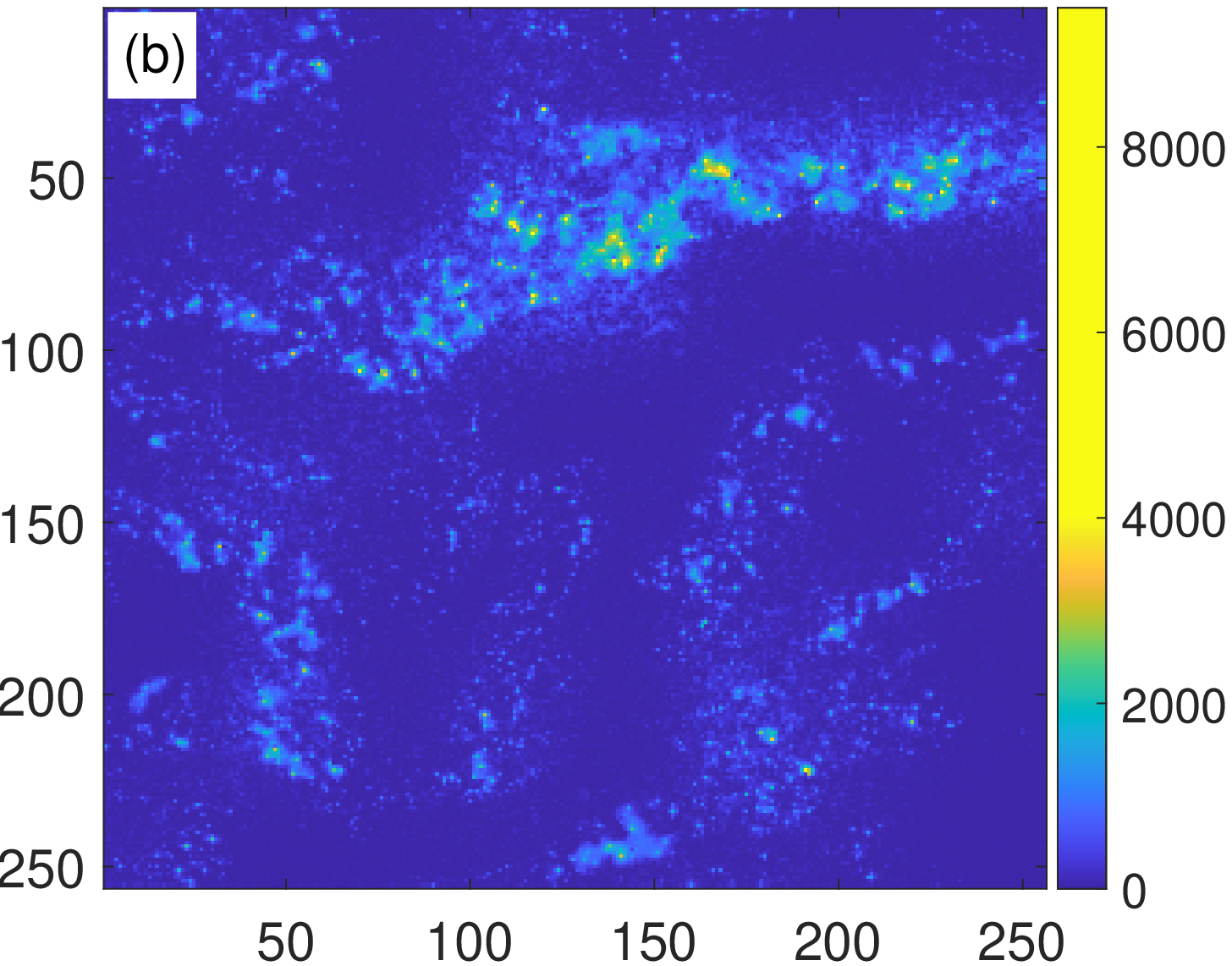}\label{fig:recons_LSC}}\\
\caption{(a) Temperature map after applying the smoothing algorithm with $n_s=5$ to block temperatures with $l_b = 32$. (b) Walker Lake data reconstruction for $p=0.85$ by using the single checkerboard implementation with site-specific temperatures (SST).}
  \label{fig:LSC}
\end{figure}

\begin{figure}[t!]
\centering
\subfigure{\includegraphics[scale=0.52,clip]{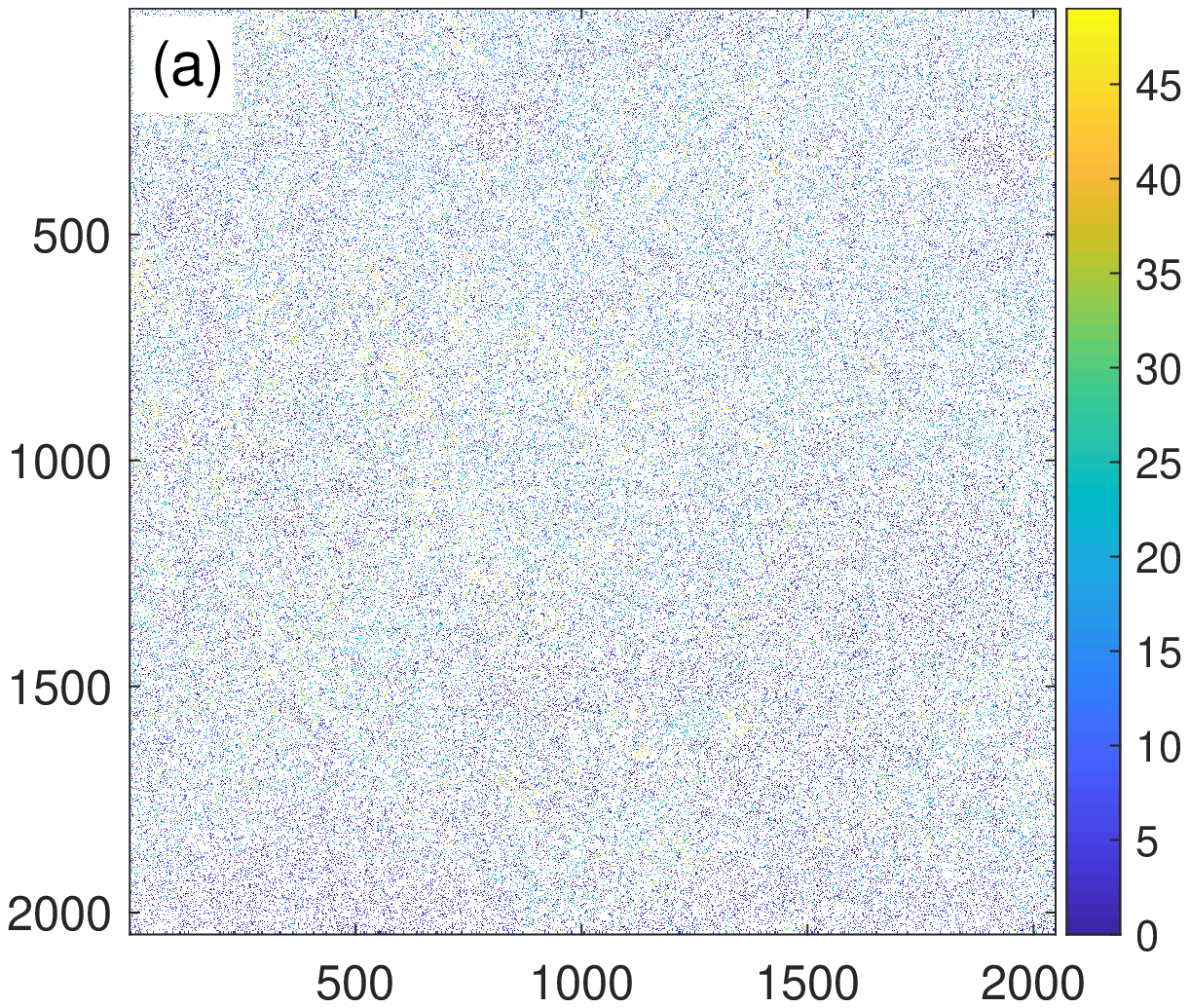}\label{fig:missing_data_p085_kaibab_plateau}}
\subfigure{\includegraphics[scale=0.52,clip]{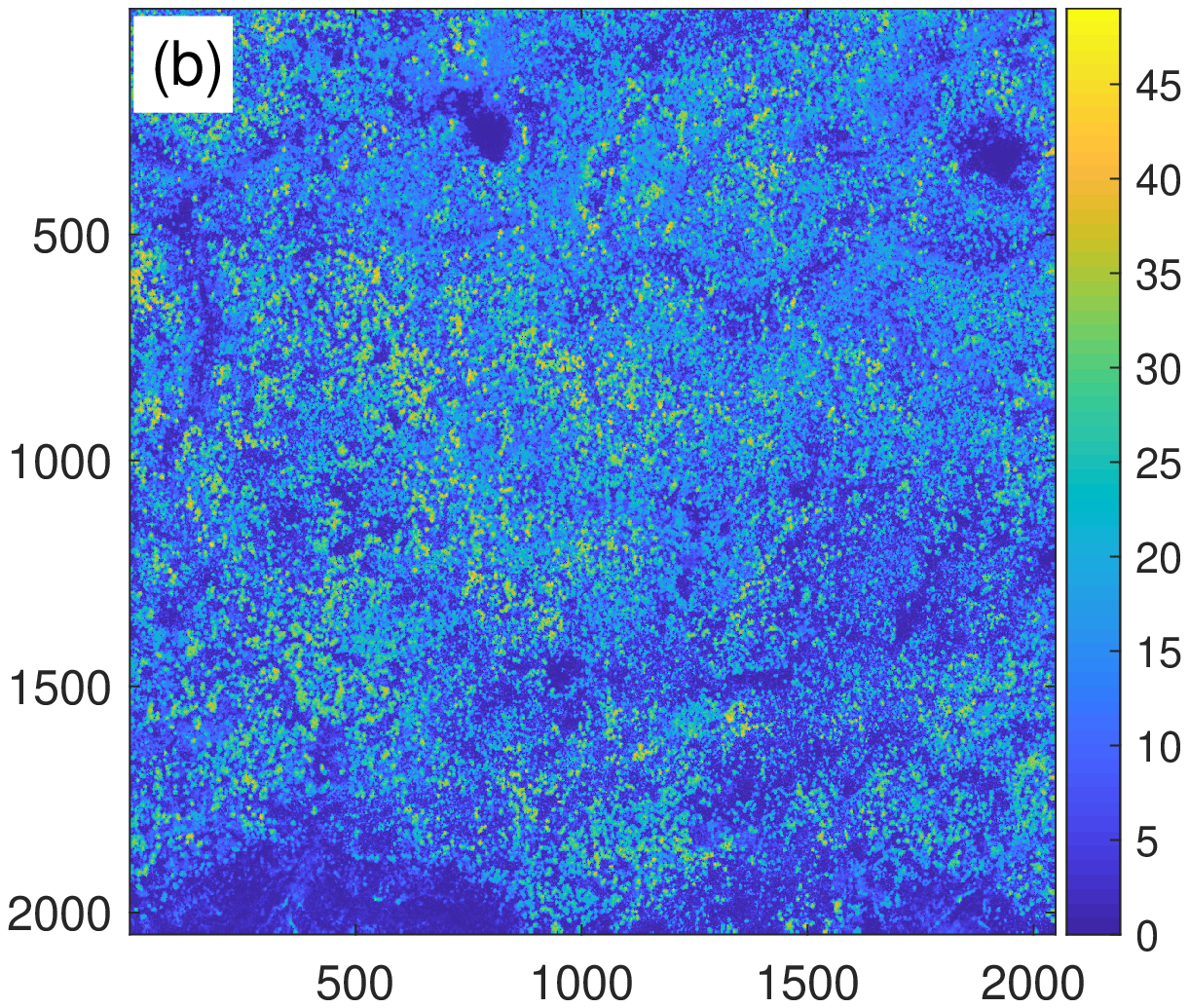}\label{fig:R_LSC_Smooth5_p085_kaibab_plateau}}
\subfigure{\includegraphics[scale=0.52,clip]{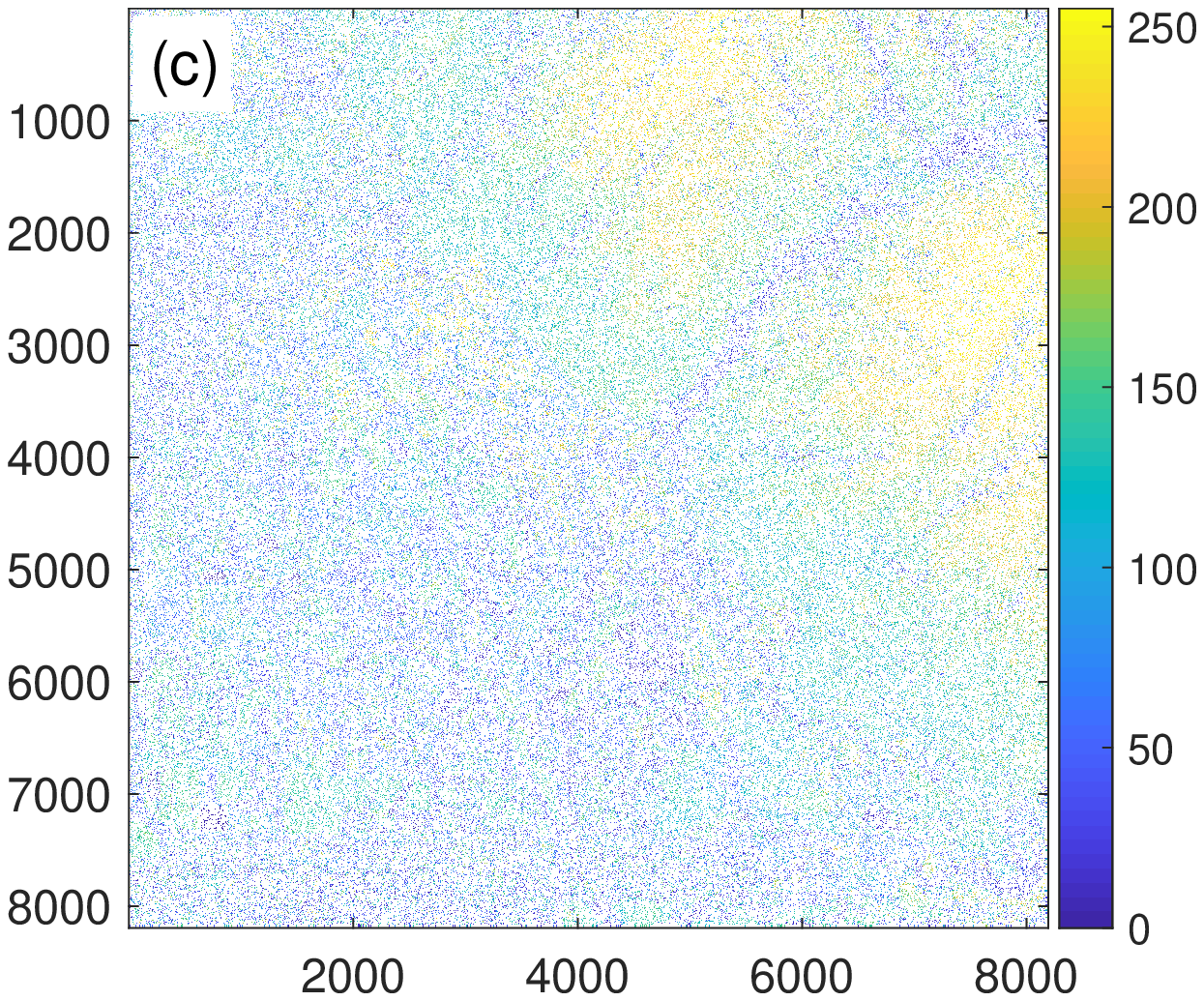}\label{fig:missing_data_p085_wasatch_front}}
\subfigure{\includegraphics[scale=0.52,clip]{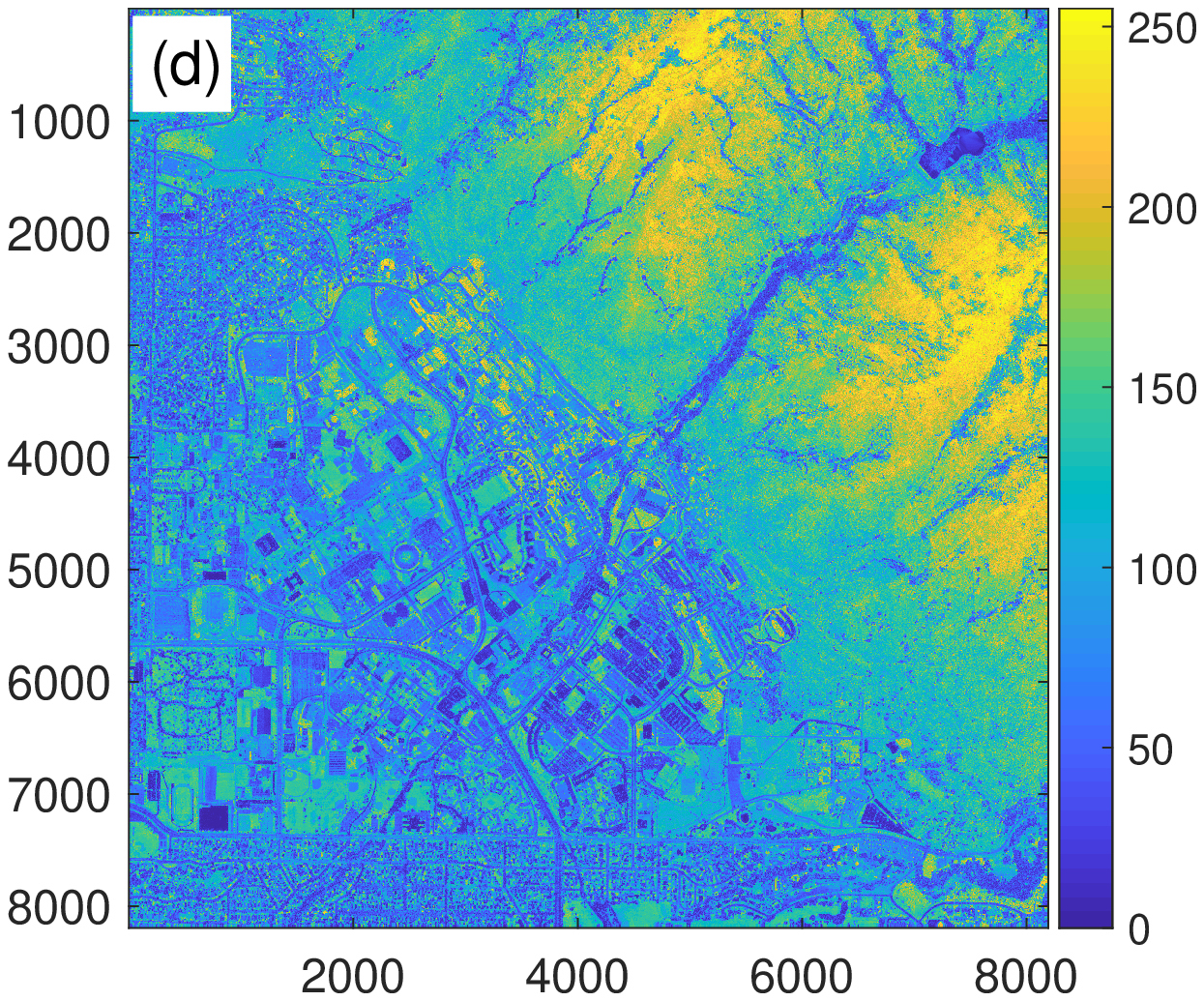}\label{fig:R_LSC_Smooth5_p085_wsatch_front}}
\caption{(a,c) Samples of the Kaibab plateau and Wasatch front, obtained by removing $p=85\%$ of pixels, and (b,d) their reconstructions obtained by the \svmpr SST method.}
\label{fig:recon}
\end{figure}

The second implementation of the \svmpr model attempts to eliminate the artifacts associated with the BST approach by applying a smoothing algorithm to the block-specific temperatures to obtain a smooth variation of the temperature values over the entire grid. Consequently, each site is assigned an individual value of the reduced temperature and thus this implementation will be referred to as the \svmpr method with site-specific temperatures (SST). The block-specific temperatures for Walker lake data with $l_b = 32$ (see Fig.~\ref{fig:block_32_temps}) after smoothing are shown in Fig.~\ref{fig:block_32_smooth_temps}. Using these local temperatures in the conditional simulations we obtain the data reconstruction for the Walker lake data, shown in Fig.~\ref{fig:recons_LSC}. Notice that the unnatural boundary effects from the BST implementation are now completely eliminated while the spatial variability in different regions is reproduced much better than in the original \mpr method (compare to Fig.~\ref{fig:recons_SC_085}). In principle in this \svmpr (SST) implementation both the standard single and the double checkerboard decompositions can be used but for simplicity, we have chosen to use the former one. Applying the \svmpr approach in Fig.~\ref{fig:recon} we also visually present the results of the reconstruction of the Kaibab plateau and Wasatch front data sets after randomly removing $p=85\%$ of pixels. 

\begin{figure}[t!]
\centering
\includegraphics[scale=0.5,clip]{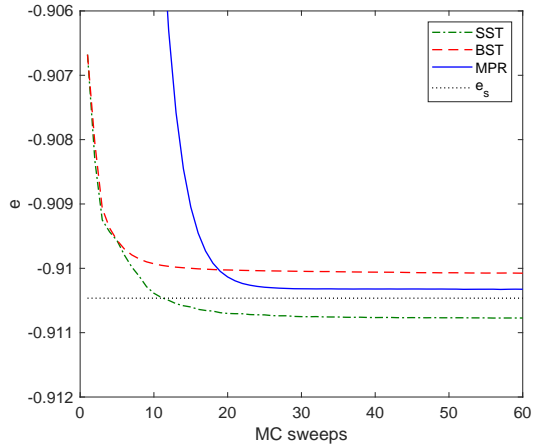}
\caption{Evolution of the specific energy during equilibration process in the respective \mpr-based methods for the Wasatch front data with $p=0.3$ averaged over 100 realizations. $e_s$ represents the average of the sample energy.}
\label{fig:Equi_energies_wasatch_p03}
\end{figure}

We note that in both the BST and the SST implementations, due to the splitting of the simulations on a large grid into a number of parallel simulations in much smaller blocks (including at most hundreds of sites), the equilibration even faster than in the original \mpr method. To demonstrate the impact of the domain spitting used in the \svmpr methods, in Fig.~\ref{fig:Equi_energies_wasatch_p03} we illustrate the equilibration process in the biggest Wasatch front data set with $p=0.3$. In particular, we show the evolution of the specific energy $e$, calculated for the whole grid, in the respective \mpr-based methods averaged over 100 realizations. One can notice that event though a random initialization in the \mpr method results in the initial value of $e$ far from the equilibrium value, the relaxation process is relatively fast even for such large data set requiring only about 30 MC sweeps. Nevertheless, the \svmpr BST and SST implementations can  shorten it even more due to the initialization by the per-block averages, corresponding to the specific energies much closer to equilibrium values. It is also interesting to compare the equilibrium energy values $e_{eq}$ resulting from different approaches among each other as well as with the sample-estimated value $e_s$ (see Eq.~\eqref{eq:mpr-sse}). One can see that while the \mpr method gives $e_{eq}$ close to $e_s$, the BST and SST implementations give the values of $e_{eq}$ respectively higher and lower than $e_s$. The increased BST value can be explained by the presence of boundaries between different blocks, which create unnatural domain walls and thus increase the total energy. On the other hand, the smoothening of the temperatures in the SST implementation leads to partial elimination of the domain walls, not only those unnaturally created by the BST approach but also those substantiated by the data.     

\subsection{Statistical validation}
\begin{figure}[t!]
\centering
\includegraphics[scale=0.47,clip]{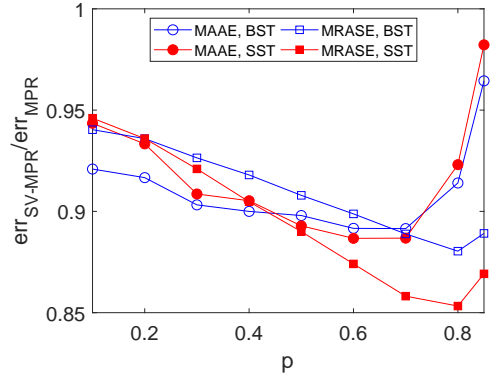}
\caption{Errors of the \svmpr implementations (BST and SST), $err_{\rm SV-MPR}$, relative to those obtained from the simple \mpr method, $err_{\rm MPR}$, as functions of the data thinning $p$, for the Walker lake data set.}
  \label{fig:err-p}
\end{figure}

The results for the Walker lake data set obtained by the standard \mpr method as well as the \svmpr BST and SST implementations are shown in Table~\ref{table:errors} in terms of the prediction errors and computational times for different degrees of the sample sparsity. As expected, both MAAE and MRASE errors increase with higher ratios of the missing data $p$ for all implementations. Nevertheless, it is clear that implementing a spatial dependence in the simulation temperature has a significant positive effect on the prediction accuracy. Compared to the \mpr approach with a single global temperature the values of both MAAE and MRASE in the \svmpr implementations are considerably smaller for all $p$. The relative improvements achieved by the present implementations are demonstrated in Fig.~\ref{fig:err-p}, in which we present error ratios of the respective methods as a function of the data thinning. Not surprisingly, the benefits of the parameter spatial variability increase with the increase of the sample sparseness, i.e., the decrease of the amount of the local conditioning data. The relative MAE (MRASE) errors decrease up to $p \approx 0.7$ ($p \approx 0.8$), where the \svmpr accuracy outperforms the \mpr one by up to 11\% (15\%), and then they start increasing. Their increase for very large $p$ is related with the above discussed artifacts of the second kind - misestimation of the local (block-specific) parameters due to insufficient statistics within the respective blocks. By comparing the BST and SST implementations of the \svmpr approach, the former appears to be more accurate for smaller $p$, while the benefits of the latter show up more at intermediate and larger $p$.
%
\begin{table}[t!]
\caption{Validation measures for the three implementations of the \mpr-based prediction algorithm for various missing data ratios $p$ applied to the Walker lake pollution data set. All implementations use single precision (FP32) arithmetic with hardware intrinsic CUDA functions, $l_b$ = 32 and for the SST implementation $n_s=5$.}
\begin{tabular}{ lccccccccc}
\hline\noalign{\smallskip}
  &\multicolumn{3}{c}{\mpr} &   \multicolumn{3}{c}{\svmpr (BST)} &   \multicolumn{3}{c}{\svmpr (SST)}\\
\hline
$ p$ & MAAE &  MRASE & t[ms] &  MAAE &  MRASE & t[ms] &  MAAE &  MRASE  &  t[ms]\\
 \hline
0.1&   $ 177 $&    $ 352 $&  $7.3$&   $ 163 $&  $ 331 $&  $6.2$&   $ 167 $&    $ 333 $&    $6.4$\\
0.2&   $ 180 $&    $ 359 $&  $7.5$&   $ 165 $&   $ 336 $&  $6.3$&    $ 168 $&    $ 336 $&    $6.9$\\
0.3&   $ 186$&    $ 367 $&  $7.5$&   $ 168 $&   $ 340$&  $6.5$&    $ 169 $&    $ 338 $&    $6.6$\\
0.4&   $ 190 $&    $ 378 $&  $7.8$&   $ 171 $&   $ 347 $&  $5.3$&    $ 172 $&    $ 342 $&    $6.3$\\
 0.5&   $ 196$&    $ 391 $&  $7.6$&   $ 176 $&   $ 355$&  $5.3$&    $ 175 $&    $ 348 $&   $5.7$\\
 0.6&   $ 203$&    $ 405 $&  $7.5$&   $ 181 $&   $ 364$&  $5.2$&    $ 180 $&    $ 354 $&   $5.4$\\
 0.7&   $ 212 $&    $ 423 $&  $7.5$&   $ 189 $&   $ 376$&  $5.2$&    $ 188 $&    $ 363 $&   $5.3$\\
 0.8&   $ 221 $&    $ 443 $&  $8.0$&   $ 202 $&   $ 390$&  $5.3$&    $ 204 $&    $ 378 $&   $5.3$\\
 0.85&   $ 225 $&    $ 451 $&  $7.5$&   $ 217 $&   $ 401$&  $5.3$&    $ 221 $&    $ 392 $& $5.4$\\
\noalign{\smallskip}\hline
\end{tabular}
\label{table:errors}
\end{table}
As already demonstrated by~\citet{MPR-GPU}, the GPU implementation of the \mpr model resulted in a computationally very efficient prediction method. In the Walker lake data set the \mpr prediction of arbitrarily large portion of missing data takes no more than 8 ms, with no apparent dependence on $p$. Both the BST and SST implementations of the \svmpr algorithm, besides the above demonstrated increasing of the prediction accuracy, also further decrease the computational burden with the computational time squeezed to 5-7 ms. The computational complexity of the \svmpr algorithm will be discussed in a more detail below.

%
%
The results for the remaining (much larger) data sets are summarized in Table~\ref{table:errors_2} for the missing data ratio $p = 0.85$. The expected degree of the improvement in the prediction performance, resulting from the introduction of spatial dependence in the parameter, can be judged from the character of the data under consideration. The Walker lake data set with the extensive areas corresponding to (almost) constant values is an example of the spatial distribution which can greatly benefit from the spatially-variable parameters. To certain extent similar features, with some larger areas of constant values, can also be observed in the remaining data sets. Therefore, the \svmpr implementations can also be expected to deliver better prediction performance than the \mpr method. Indeed, the \svmpr implementations reduce the \mpr MAAE errors by 4-11\% (BST) and 0.4-7.4\% (SST) and the MRASE errors by 7.4-8.5\% (BST) and 5.3-6.3\% (SST). \\
\begin{table}[t!]
\caption{Validation measures for the three implementations of the \mpr-based prediction algorithm for the Kaibab plateau and Wasatch from data sets with missing data ratio $p = 0.85$. The implementations use single precision (FP32) arithmetic with hardware intrinsic CUDA functions. The linear sizes of these data sets are: $L = 2\ 048$ for the Kaibab plateau and $L = 8\ 192$ for the Wasatch front data sets.}
\begin{tabular}{ lccccccccc}
\hline\noalign{\smallskip}
  &\multicolumn{3}{c}{\mpr} &   \multicolumn{3}{c}{\svmpr (BST)} &   \multicolumn{3}{c}{\svmpr (SST)}\\
\hline
 Dataset &  MAAE &  MRASE & t[s] &  MAAE &  MRASE & t[s] &  MAAE &  MRASE  &  t[s]\\
 \hline
  Kaibab plateau &   $ 5.00$&    $ 7.05$& $0.175$&  $ 4.81 $&   $ 6.53$& $0.161$&   $ 4.98 $&    $ 6.68 $& $0.193$\\
  Wasatch front &   $ 26.14$&    $ 35.60$& $2.64$&  $ 23.36 $&   $ 32.58$& $2.44$&   $ 24.22 $&    $ 33.37 $& $3.02$\\
\hline\noalign{\smallskip}
\end{tabular}
\label{table:errors_2}
\end{table}
\indent As for the computational efficiency, some comments are in order. Compared to the original \mpr approach, one can notice overall only a small reduction of $t$ (if any) achieved by the BST implementation and even some increase by applying the SST implementation. This might appear as contradiction with the above results for Walker lake data, presented in Table~\ref{table:errors}, but it has a simple explanation. The total execution time largely depends on the speed of the convergence to equilibrium, as the rest is only dictated by the number of samples needed for averaging and is almost independent of the chosen method. However, as explained above and demonstrated in Fig.~\ref{fig:Equi_energies_wasatch_p03}, the \svmpr implementations are indeed either faster in reaching the equilibrium state (BST) or more efficient in finding states with the energy levels lower than those achievable by the \mpr method (SST). The reason why the shorter BST equilibration time is not more apparently reflected in the values of $t$ is that, for simplicity, in all the methods we used the same (default) values of the parameter $n_{\rm fit}=20$, which defines the memory length of the energy time series employed in testing the onset of equilibrium, and the parameter $n_f=5$, which defines the frequency of verification of equilibrium conditions\footnote{For detailed decription of $n_{fit}$ and $n_f$ see also paper by~\cite{MPR-GPU}.}. Thus, in all the used approaches the first checking of the equilibrium conditions is only performed after the first $n_{\rm fit} + n_f = 25$ MC sweeps, regardless of the fact that in some cases the equilibrium is reached much faster. For example, from Fig.~\ref{fig:Equi_energies_wasatch_p03} we can see that the equilibrium conditions of the BST implementation are already reached at about 20 MC sweeps even for as big data as Wasatch front. On the other hand, the SST implementation requires more (about 25) MC sweeps but it reaches lower energy levels. Additional contribution, which further increases the total SST execution time, comes from the temperature smoothing procedure. Thus, the execution time of the \svmpr implementations can be in principle shortened by resetting of the concerned parameters. However, considering the superior efficiency of all the \mpr-based methods compared to some other approaches (e.g., see comparison with IDW in Fig.~\ref{fig:time_mpr_idw}), at present we find it unessential.


\subsection{Effect of \svmpr parameters}
\begin{figure}[t!]
\centering
\subfigure{\includegraphics[scale=0.47,clip]{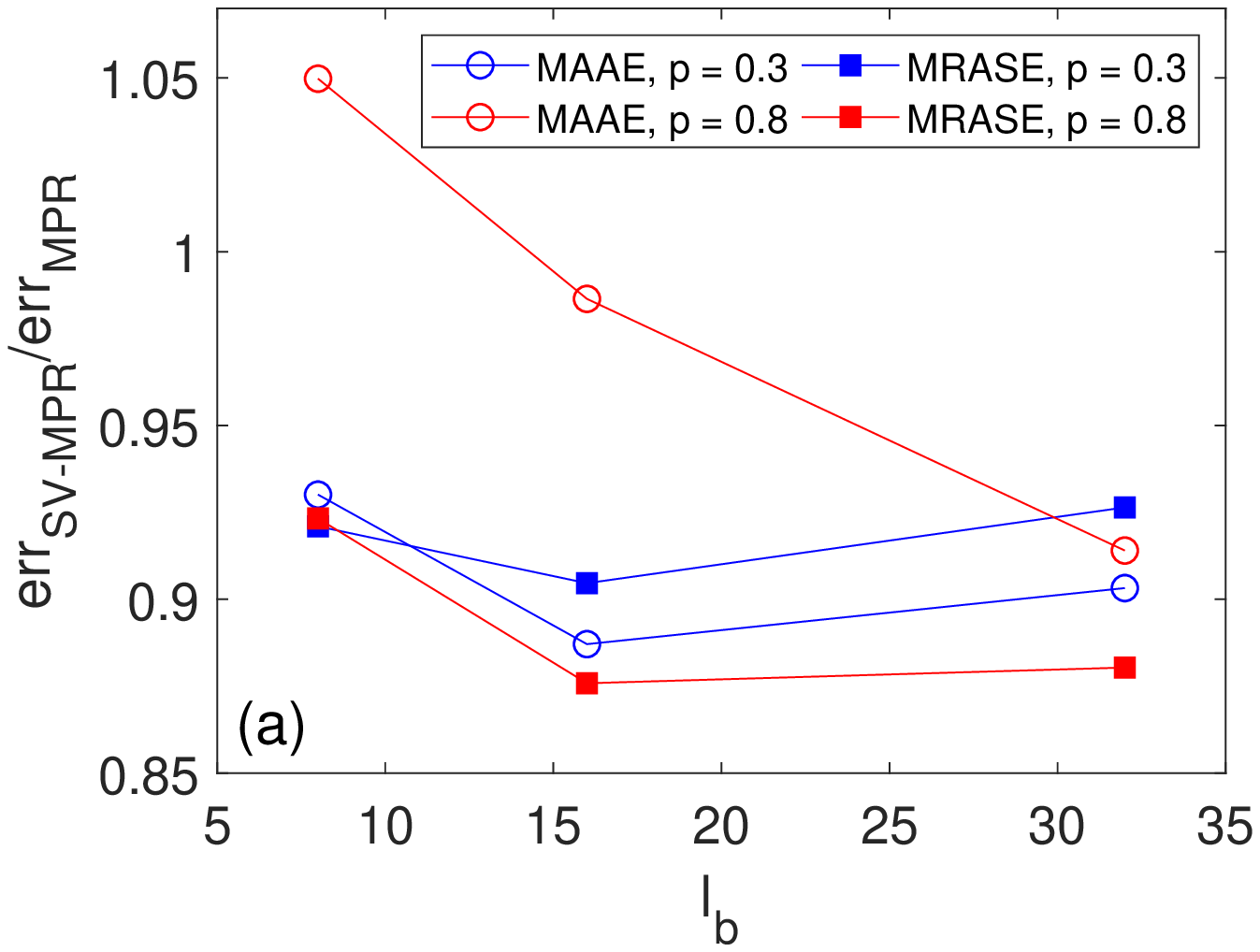}\label{fig:err-Nb}}
\subfigure{\includegraphics[scale=0.47,clip]{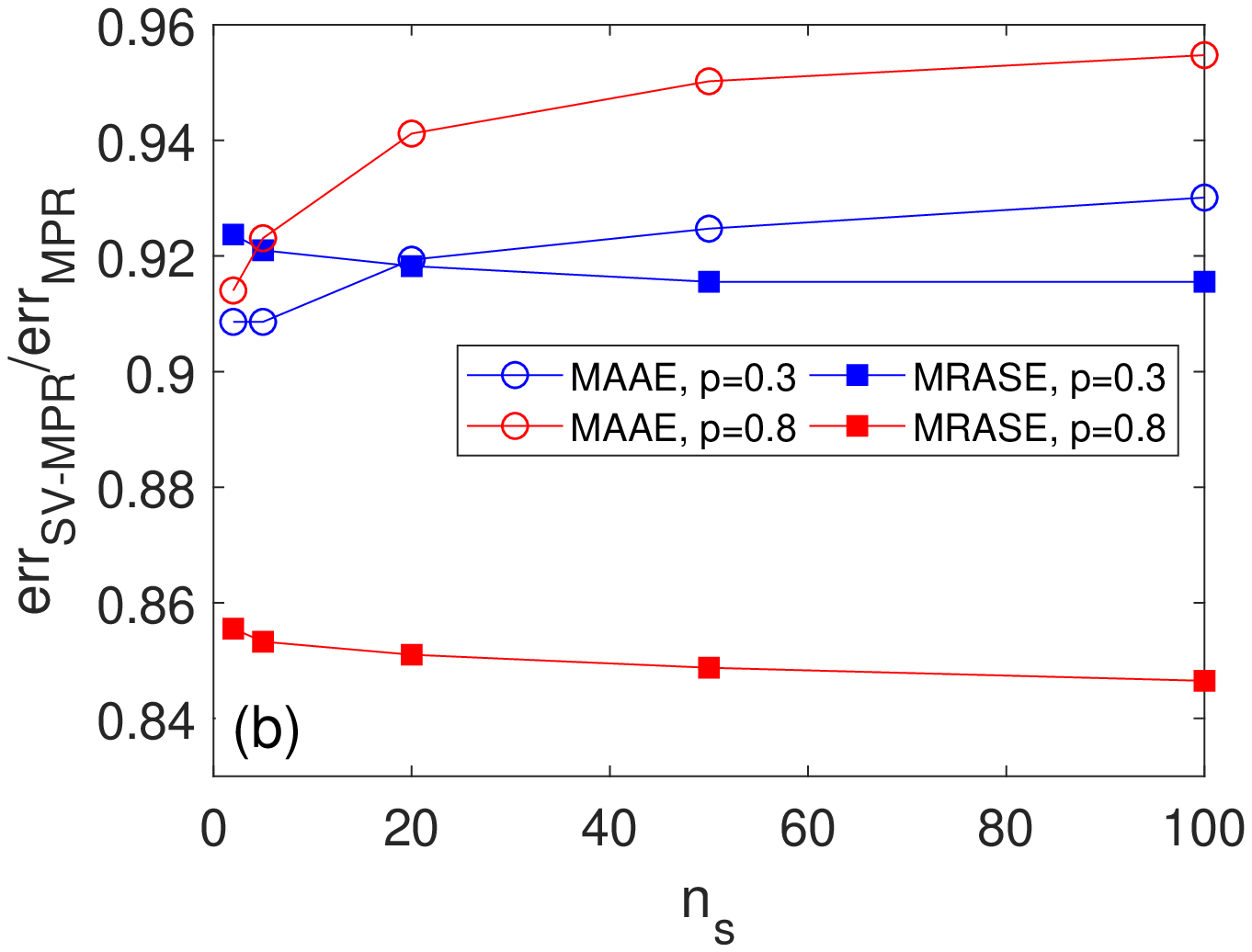}\label{fig:err-smooth}}
\caption{(a) Errors of the \svmpr method, $err_{\rm SV-MPR}$, relative to those obtained from the simple \mpr method, $err_{\rm MPR}$, as functions of (a) the block size $l_b$ in the BST implementation and (b) the smoothing parameter $n_s$ in the SST implementation.}
  \label{fig:err}
\end{figure}

In the above study we used the \svmpr methods with the fixed block size $l_b=32$ and in the SST implementation the fixed smoothing parameter $n_s=5$. In the following we demonstrate the effect of these parameters on the prediction performance of the respective methods. The latter is demonstrated in Fig.~\ref{fig:err}, in which the prediction errors of the \svmpr method, $err_{\rm SV-MPR}$, are presented relative to those obtained from the simple \mpr method, $err_{\rm MPR}$ for two cases of a relatively small ($p=0.3$) and large ($p=0.8$) degrees of thinning. In Fig.~\ref{fig:err-Nb} they are shown for the BST implementation considering different block sizes $l_b$ and in Fig.~\ref{fig:err-smooth} for the SST implementation with the fixed $l_b=32$ and a varying parameter $n_s$. 
As one can see from Fig.~\ref{fig:err-Nb}, the above used block size $l_b=32$ may not be optimal in terms of minimizing prediction errors. As discussed in the previous section, its decreasing allows a greater flexibility in capturing the local variability, which may result in improvement of the prediction performance. On the other hand, smaller values of $l_b$ in combination with larger values of $p$ suffer from the lack of the sampling points within the blocks, which may lead to the block-specific parameters misestimation and consequently deterioration of the prediction accuracy. 

Figure~\ref{fig:err-smooth} shows that the smoothing parameter $n_s$ not only suppresses the edge-like visual artifacts but its choice can also affect the prediction performance of the SST implementation. In particular, the gradual smoothing tends to increase MAAE and decrease (to smaller extent) MRASE and its effect is more pronounced in data with larger sparsity. Therefore, relatively small values of the smoothing parameter, such as $n_s \approx 5$, which to some degree suppress the visual artifacts but still do not excessively increase MAAE, may be considered as an acceptable compromise. In any case, compared to the standard \mpr method, the SST implementation of the \svmpr method appears to deliver superior prediction performance for arbitrary choice of the parameter $n_s$.

\subsection{Comparison with established IDW approach}
Prediction performance and computational efficiency of the original \mpr method have been compared with several established interpolation methods~\citep{mz-dth18}. Among them, the inverse distance weighted (IDW)~\citep{shepard1968two} and the ordinary kriging (OK)~\cite{wack03} were the methods that gave the prediction performance comparable with the \mpr approach. However, the high computational complexity of OK prevents it to be applied to huge data sets. Therefore, for comparison of the present GPU-implemented \mpr-based methods we chose the GPU-implemented IDW method~\citep{marcel17} by using the CUDA code available at GitHub~\citep{ruggieri2017}. The implementation of IDW involves choosing two parameters, which can influence both the prediction performance and computational efficiency. 
The power parameter was set to a default value of 2 and the search radius $R$ was varied from the minimum for which every prediction point has some sample points within the radius up to the maximum involving practically all points on the grid. We note that an optimal choice of $R$ is not so obvious. It can be set to some fixed value equal for all points, which can result in the problem of some prediction points not having any samples in the search radius, or it can be made variable, which would make the IDW implementation more involved.
%
\begin{figure}[t!]
\centering
\subfigure{\includegraphics[scale=0.47,clip]{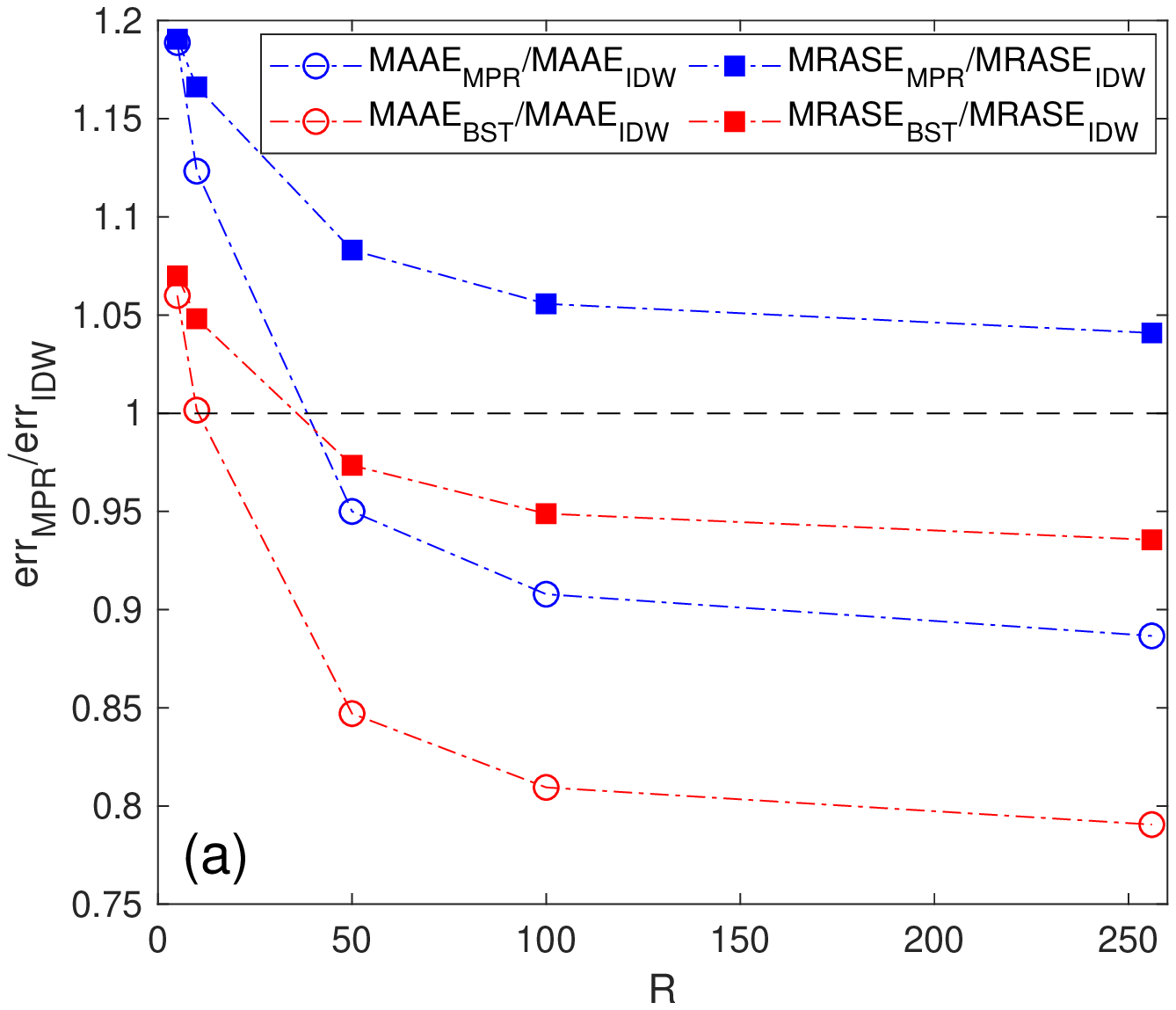}\label{fig:idw_walker_p06_new}}
\subfigure{\includegraphics[scale=0.47,clip]{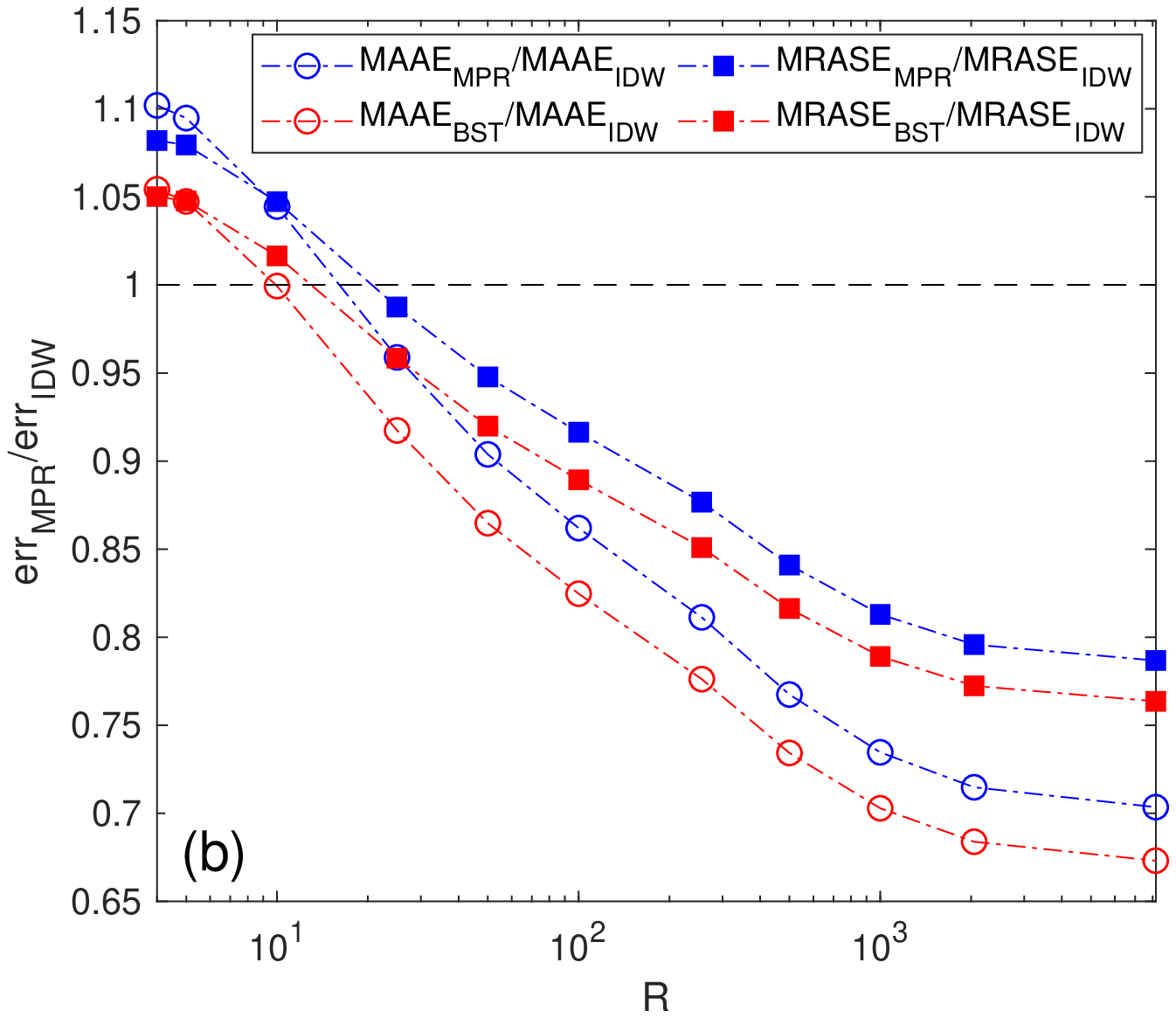}\label{fig:idw_wasatch_p06_new}}
\caption{MAAE (open circles) and MRASE (filled squares) errors of the \mpr (blue color) and \svmpr BST (red color) methods, relative to those obtained from the IDW method, as functions of the search radius $R$, for (a) Walker lake and (b) Wasatch front
data sets.}
  \label{fig:err_mpr_idw}
\end{figure}
In particular, in Fig.~\ref{fig:err_mpr_idw} we present errors of the \mpr (blue color) and \svmpr BST (red color) methods, relative to those obtained from the IDW method, $err_{\rm MPR}/err_{\rm IDW}$ and $err_{\rm BST}/err_{\rm IDW}$, as functions of the search radius $R$, for (a) Walker lake and (b) Wasatch front
data sets. Therefore, the values larger than 1 mean superior performance of IDW. One can notice that IDW shows the best prediction performance for the smallest search radius $R$, where it 
outperforms both \mpr as well as \svmpr BST methods. However, with the increasing $R$ the IDW errors increase and in most cases (except MAAE error for Walker Lake data) starting from some value of $R$ both \mpr-based methods become superior. Owing to the improved prediction performance of the present \svmpr BST implementation, the latter outperforms IDW starting from much smaller $R$ than the standard \mpr method. Even for the optimal IDW performance at small $R$, the errors of the \svmpr BST method do not exceed those from IDW by more than $5-7$\%
. 

\begin{figure}[t!]
\centering
\subfigure{\includegraphics[scale=0.47,clip]{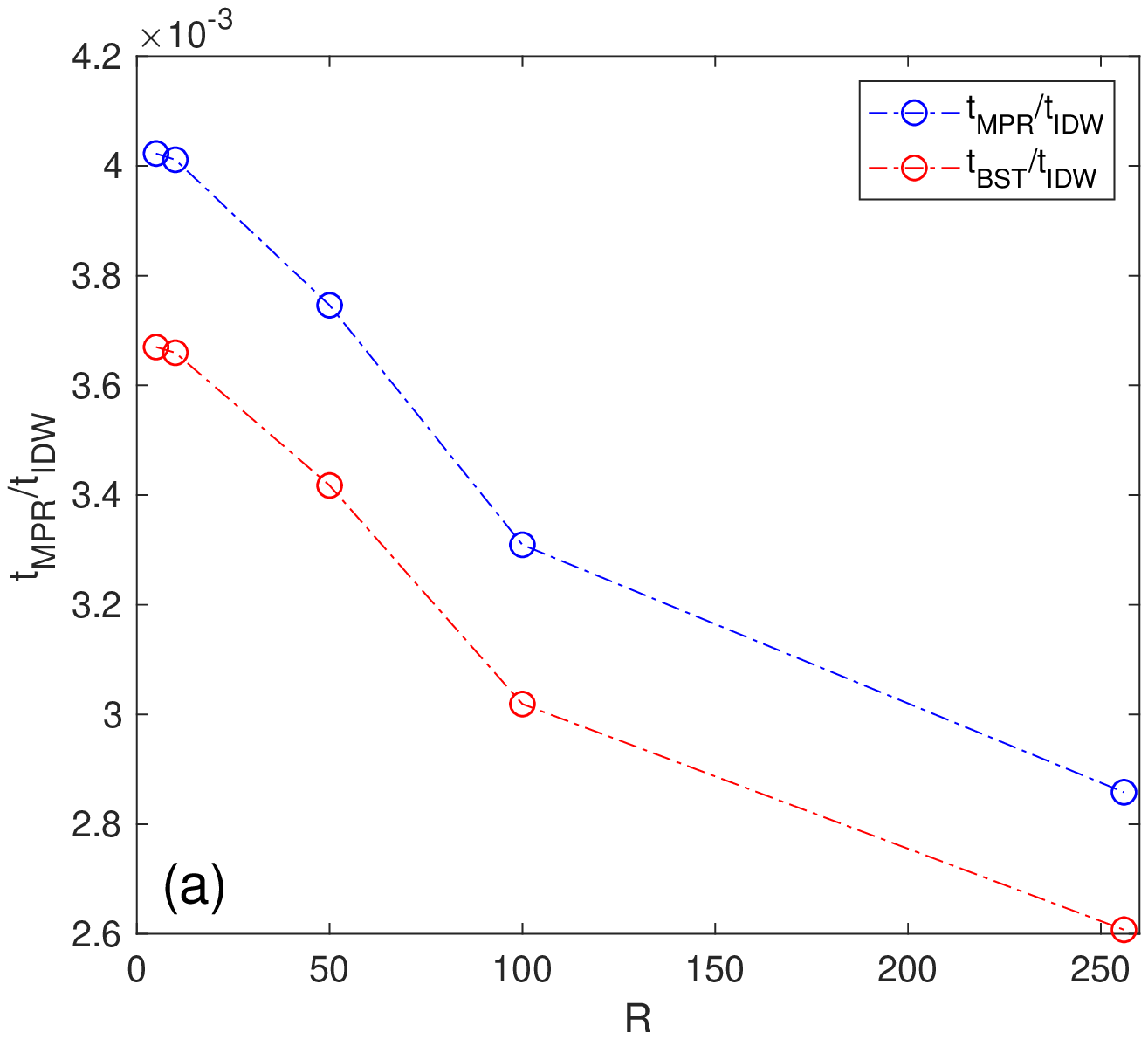}\label{fig:time_idw_walker_p06_new}}
\subfigure{\includegraphics[scale=0.47,clip]{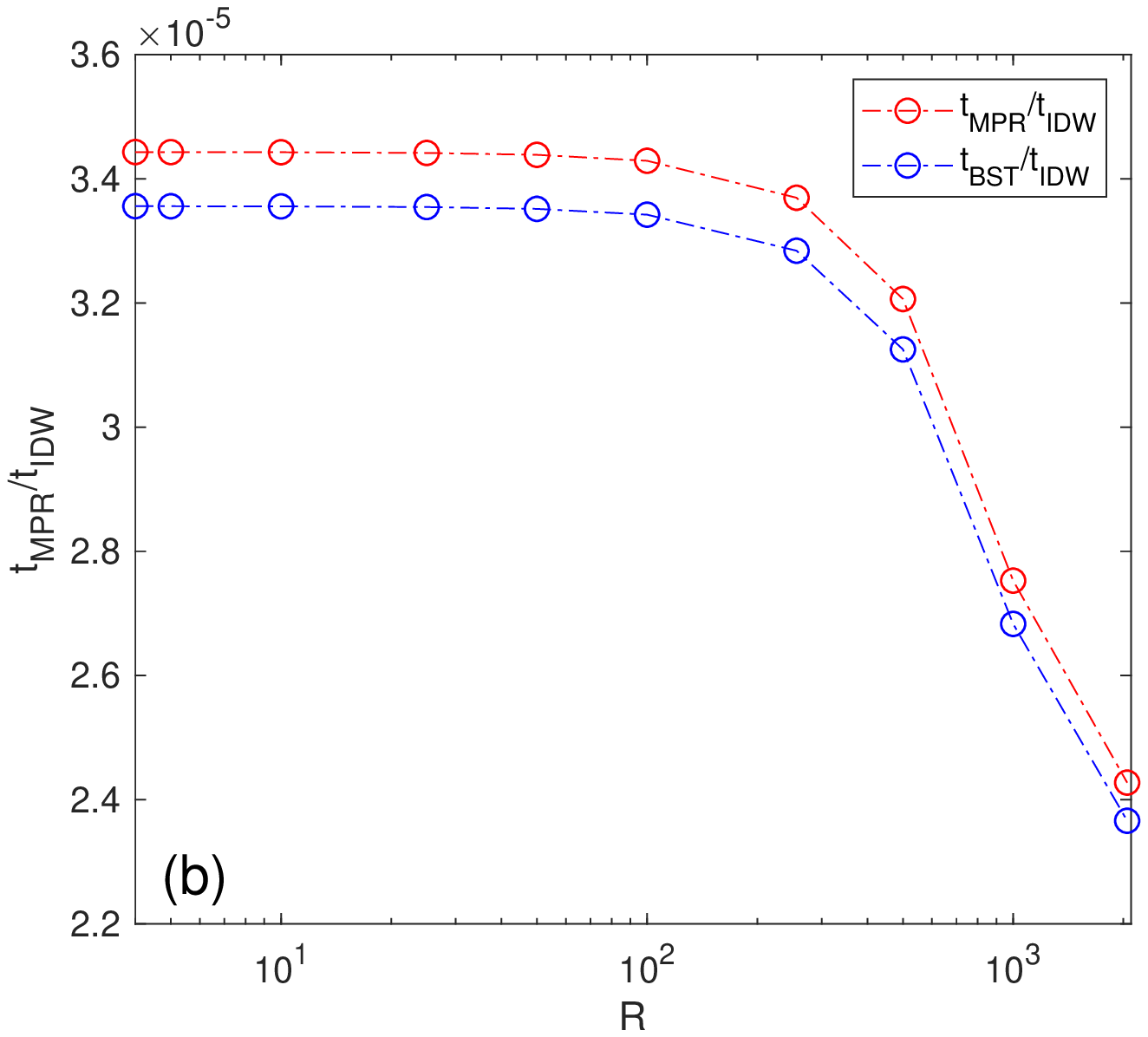}\label{fig:time_idw_wall3_p06_new}}
\caption{GPU time of the \mpr (blue color) and \svmpr BST (red color) methods, relative to those required by the IDW method, as functions of the IDW search radius $R$, for (a) the Walker lake and (b) Kaibab plateau data sets.}
  \label{fig:time_mpr_idw}
\end{figure}

On the other hand, the computational efficiency of the \mpr-based methods clearly dominates over the IDW performance, regardless of the parameters. Due to favorable scaling properties of the \mpr method (computational time scales approximately linearly with system size) its dominance over IDW further increases with the data size. As presented in Fig.~\ref{fig:time_mpr_idw}, the ratio of the GPU execution times of the \mpr-based to IDW methods is of the order of $10^{-3}$ for relatively small Walker lake data set ($L=256$) and of the order of $10^{-5}$ for much larger Kaibab plateau data set ($L=2\ 048$). We note that for as massive data as Wasatch front ($L=8\ 192$) the IDW calculations could not be executed on a standard GPU-equipped PC within some reasonable time at all\footnote{We terminated the calculation after about two weeks of running and the errors presented in Fig.~\ref{fig:idw_wasatch_p06_new} were obtained by executing the code on the supercomputer Govorun at the Joint Institute for Nuclear Research in Dubna.}. The curves presented in Fig.~\ref{fig:time_mpr_idw} also show that the relative computational efficiency of the \mpr-based methods, es expected, increases with the search radius $R$ due to decreasing efficiency of IDW. Furthermore, by comparing the results for the two \mpr based approaches one can see that the \svmpr BST method is somewhat faster than the original \mpr approach, again in line with the expectations discussed above.

\section{Conclusions}
\label{sec:conclusion}
In the current study we proposed a rather general approach to modelling spatial heterogeneity, the feature often present in massive spatial data, in GPU-implemented spatial prediction methods for gridded data. In particular, we presented two approaches to introducing spatial dependence to the model parameter (temperature) by the so-called double-checkerboard (DC) decomposition to our previously introduced GPU-accelerated \mpr method and thus obtained two \svmpr methods with spatially varying temperatures. In the BST variation, separate values of the temperature are obtained for each block of the DC-decomposed grid and in the SST variation even each individual prediction point is modelled using its own temperature. Then, similar to the \mpr method, predictions of unknown values are obtained from conditional situations. However, in the present methods the conditional situations are actually non-equilibrium but assume ``local'' equilibrium conditions corresponding to the local temperatures. Using various types of big heterogeneous real data, such as remote sensing data
, we have demonstrated that the proposed \svmpr methods significantly improve prediction performance and even computational efficiency of the original \mpr method. Their prediction performance is competitive with some established prediction methods, such as IDW, but their execution times are by several orders of magnitude faster. We note that the presented approach to modelling spatial heterogeneity was demonstrated on the \mpr method but in fact it is rather general and its application to other GPU-implemented methods is rather straightforward.
 
Future extensions of the presented models to further increase their flexibility and ability to capture various relevant features present in real data, such as geometric anisotropy or non-Gaussianity, may involve adding more parameters to the temperature, while still keeping their local nature (within blocks) and spatial variability. For example, the geometric anisotropy could be introduced by distinguishing the exchange interactions in different directions, i.e., by introducing a directional exchange interaction anisotropy parameter. The non-Gaussianity could be incorporated by including higher-order interactions and/or applying some suitable form of an external “magnetic” (bias) field to the Hamiltonian. Another possibility to be considered is inclusion of further-neighbor pair-wise interactions that would, for example, control data smoothness. The impressive computational efficiency of the models offers possibility to extend them to three dimensions (3D space or 2D space + time), where there is still lack of efficient methods that would enable modeling massive data~\citep{Wang_2012}. The generalization of the GPU code from two to three dimensions is straightforward. The GPU time in 3D is expected to increase by the factor of $3/2$ but the relative efficiency of the CPU and GPU implementations should be preserved~\citep{weig12}.

\section*{CRediT authorship contribution statement}
{\bf Mat\'{u}\v{s} Lach:} Software, Investigation, Data curation, Visualization, Validation, Writing -- original draft preparation. {\bf Milan \v{Z}ukovi\v{c}}: Conceptualization, Methodology, Formal analysis, Validation, Writing -- review and editing, Supervision. All authors contributed to the writing of the manuscript.

\section*{Declaration of competing interest}
The authors declare that they have no known competing financial interests or personal relationships that could have appeared to influence the work reported in this paper.

\section*{Acknowledgements}
\label{sec:acknowledgements}
This work was supported by the Scientific Grant Agency of Ministry of Education of Slovak Republic (Grant No. 1/0531/19) and the Slovak Research and Development Agency (Grant No. APVV-20-0150).

\section*{Computer code availability}
\noindent Name of the code/library: packages MPR, $\mathrm{SV-MPR\ (BST)}$, and $\mathrm{SV-MPR\ (SST)}$ \\
Contact: e-mail and phone number: matus.lach@gmail.com, +421 915 609 735 \\
Hardware requirements: NVIDIA GPU with compute capability 5.2 or higher (GTX 950 or newer) \\
Program language: CUDA \\
Software required: CUDA 10.1, CUB library, Windows or Linux operating system (not tested on OS X) \\
Program size in zipped format: 5 MB for MPR, 3.3 MB for $\mathrm{SV-MPR\ (BST)}$, and 9.4 MB for 
The source codes are available for downloading at the links: \url{https://github.com/MatusLach/MPR}; \url{https://github.com/MatusLach/SV-MPR_BST}; \url{https://github.com/MatusLach/SV-MPR_SST}

%


\bibliographystyle{cas-model2-names}

\bibliography{bibliography}





\end{document}